\documentclass[a4paper]{article}

\usepackage{fullpage}

\usepackage{graphicx}

\usepackage[version=3]{mhchem}

\usepackage{amsmath} 

\usepackage{amsfonts}

\DeclareMathOperator{\e}{e}

\usepackage{textcomp}

\date{}

\usepackage{multicol}

\author{Denis Michel\\
       Universite de Rennes1-IRSET. Campus de Beaulieu Bat. 13. 35042 Rennes cedex\\
 \ denis.michel@live.fr}

\title{Kinetic approaches to lactose operon induction and bimodality}

\begin{document}
\maketitle
\textbf{Abstract}\\
\newline
The quasi-equilibrium approximation is acceptable when molecular interactions are fast enough compared to circuit dynamics, but is no longer allowed when cellular activities are governed by rare events. A typical example is the lactose operon (\textit{lac}), one of the most famous paradigms of transcription regulation, for which several theories still coexist to describe its behaviors. The \textit{lac} system is generally analyzed by using equilibrium constants, contradicting single-event hypotheses long suggested by Novick and Weiner (Novick and Weiner 1957 Proc. Natl. Acad. Sci. USA. 43, 553-566) and recently refined in the study of (Choi et al. 2008 Science 322, 442-446). In the present report, a \textit{lac} repressor (LacI)-mediated DNA immunoprecipitation experiment reveals that the natural LacI-\textit{lac} DNA complex built in vivo is extremely tight and long-lived compared to the time scale of \textit{lac} expression dynamics, which could functionally disconnect the abortive expression bursts and forbid using the standard quasi-equilibrium modes of bistability. As alternatives, purely kinetic mechanisms are examined for their capacity to restrict induction through: (i) widely scattered derepression related to the arrival time variance of a predominantly backward asymmetric random walk and (ii) an induction threshold arising in a single window of derepression without recourse to nonlinear multimeric binding and Hill functions. Considering the complete disengagement of the \textit{lac} repressor from the \textit{lac} promoter as the probabilistic consequence of a transient stepwise mechanism, is sufficient to explain the sigmoidal \textit{lac} responses as functions of time and of inducer concentration. This sigmoidal shape can be misleadingly interpreted as a phenomenon of equilibrium cooperativity classically used to explain bistability, but which has been reported to be weak in this system.
\newline
\textbf{Highlights}
\begin{itemize}\setlength{\itemsep}{0.5mm}
  \item New models are presented to explain the all-or-nothing \textit{lac} phenotypes.
  \item \textit{Lac} repressor dissociation can be modeled as the final event of a Markovian chain.
  \item Kinetic origin of an induction threshold in a single repressor rebinding interval.
  \item \textit{Lac} behaviors do not require multimeric cooperativity.\\
\end{itemize}
\begin{center}
\includegraphics[width=16cm]{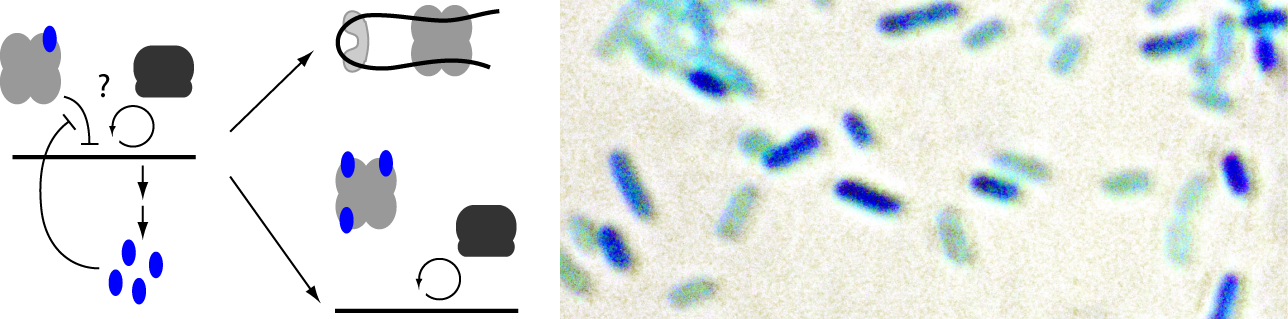} \\
\end{center}
\newpage
\begin{multicols}{2}
\section{Introduction}
Although the living cell is an open system out of equilibrium, the quasi-equilibrium approximation of fast nondriven interactions is accepted for modeling biochemical circuits using only time-independent equilibrium binding constants. Moreover, through time scale separation, stochastic interactions can generate graded outputs even in single cells (Michel, 2009). But this widespread approach no longer holds for slow and bursty molecular phenomena (Choi et al., 2010) which facilitate the crossing of switching thresholds. This could be the case for the switch of the lactose operon (\textit{lac}) of \textit{Escherichia coli} which has been proposed to result from bursting expression caused by slow binding cycles of the \textit{lac} repressor (LacI) to the \textit{lac} promoter (Choi et al., 2010). The particular situation examined here is that the switch can, or cannot, change state in a single interval between the dismantlement and the possible reformation of the \textit{lac} repression complex. This possibility is based on the experimental observation that this molecular complex naturally built in vivo, can be very long-lived at the time scale considered, thus suggesting a new way to model the \textit{lac} system. \textit{Lac} has become a celebrated model of gene regulation and a masterpiece of biochemistry textbooks, for which different interpretations exist depending on the relative kinetics of \textit{lac} expression and of the interaction between \textit{lac} DNA and LacI. Since its discovery fifty years ago (Jacob, 2011), \textit{lac} continues to attract attention as an archetype of nongenetic phenotypic switch, characterized by seemingly inconsistent behaviors depending on whether they are examined at the population or single-cell levels (Vilar et al., 2003). Upon addition of a gratuitous inducer, individual bacteria suddenly and asynchronously jump to the induced phenotype, contrasting with the progressive induction of the populational response which depends sigmoidally on both time and inducer concentration. Before identifying the molecular actors underlying these phenomena, \textit{lac} induction in the population has been interpreted as the average of many abrupt single cell switches (Benzer, 1953; Novick and Weiner, 1957). \textit{Lac} expression is regulated by inhibition of an inhibition, which is a typical positive circuit whose capacity to generate multiple steady states has long been established (Jacob and Monod, 1963; Thomas, 1998). LacI can shut off \textit{lac} expression by preventing RNA polymerase (RNAP) to either access the \textit{lac} promoter (P\textit{lac}) or to leave it (Lee and Goldfarb, 1991; Sanchez et al., 2011), with the same consequence of inhibiting reiterative transcription. The admitted mode of action of the inducer is to sequester LacI in a form no longer capable of binding the \textit{lac} operator (LacO DNA), thereby allowing RNAP to initiate transcription (Wilson et al., 2007). The induced phenotype can survive periods of absence of the inducer. This phenomenon has been attributed to an indirect feedback in which synthesis of permease, a product of the \textit{lac} operon, leads to the import of more inducer, thus clicking the system at maximal expression (Wilson et al., 2007). After LacI release from P\textit{lac}, \textit{lac} expression may produce enough permease for maintaining the bacterium in the induced state. Even in case of transient inducer withdrawal, newly supplied inducer would massively enter the cell through presynthesized permeases. Following this pioneer example of \textit{lac}, positive circuits then strikingly accumulated in the literature, suggesting that they are pivotal elements structuring gene regulatory networks (GRNs), responsible for bistability, hysteresis and memory (Thomas, 1998; Nicol-beno\^{i}t et al., 2012). However, \textit{lac} retains some distinctive features in systems biology. In their visionary study, Novick and Weiner postulated that the \textit{lac} switch could be triggered by a single random event (Novick and Weiner, 1957). This hypothesis has been completed by Choi et al. (Choi et al., 2008), who proposed that  given the low concentration of LacI in the cell, the time necessary for rebinding to P\textit{lac} can be long enough to allow \textit{lac} induction. This view in which the dissociation of LacI from LacO can be considered as a single-event, contrasts with the classical treatment with dynamic interactions, illustrated below by that of (Ozbudak et al., 2004). In this latter treatment, the equilibrium constant of LacI-LacO interaction is used in the \textit{lac} production function, implicitly assuming that both association and dissociation events between LacI and LacO are frequent compared to the synthesis of \textit{lac} products. The poor representation of single event models in the literature could be explained by the predominant belief that interactions are highly dynamic in the cell. But in fact single event models restore the initial, static view of gene repression in bacteria. In addition, it will be shown that single event mechanisms are capable of giving rise to persistent binary phenotypes, as the classical cooperative model. This possibility can solve the apparent paradox that strong equilibrium cooperativity is postulated (Ozbudak et al., 2004), whereas allosteric and cooperative phenomena were shown to be absent or very weak in the \textit{lac} system (Barkley et al., 1975; Brenowitz et al., 1991; Chen et al., 1994). These allosteric interpretations could have been misleadingly suggested by the apparent sigmoidal behaviours of \textit{lac} expression at the population level, but the brutal phenotype switch of every bacterium has been shown unrelated to population curves (Vilar et al., 2003). The fraction of induced bacteria in the population gradually increases in a sigmoidal manner, both along time for a given concentration of inducer and, symmetrically, as function of inducer concentration for a given time. Sigmoidicity in time, illustrated for example by the elegant figure obtained at low inducer concentration by Novick and Weiner (Novick and Weiner, 1957), is classically observed for positive feedbacks with reiterative amplification cycles, but this cannot be the correct explanation for \textit{lac} switches which abruptly shift from zero to max. At the population level, the \textit{lac} induction curve follows a sigmoidal function of inducer concentration, evoking equilibrium cooperative phenomena, such as:  (\textbf{i}) the cooperative fixation of the inducer in the different protomers of LacI multimers (Ozbudak et al., 2004; Zhan et al., 2010), or (\textbf{ii}) the cooperative fixation of the LacI tetramer on two operators in P\textit{lac} (Oehler et al., 1990; Oehler et al., 2006; Daber et al., 2009). This cooperativity is in turn believed to be necessary for \textit{lac} bistability. Accordingly, Hill functions proved very helpful and are extensively used in most mathematical analyses of bistability (Cherry and Adler, 2000; Sobie, 2011). But since these treatments are poorly compatible with: (\textbf{i}) the reported weakness of allosteric cooperativity in this system (Barkley et al., 1975; Brenowitz et al., 1991; Chen et al., 1994) and (\textbf{ii}) the single event hypothesis (Novick and Weiner, 1957; Choi et al., 2008; Choi et al., 2010), alternative interpretations will be provided in the present article, in an attempt to reconcile the data obtained at the population and the single cell levels. This study is restricted to the minimal core \textit{lac} regulation network, that will be shown sufficient for capturing certain characteristics of \textit{lac}.
\section{Graded \textit{vs} binary repression by LacI}
The basic differences between quasi-equilibrium and transient mechanisms are schematized in the case of the competition between RNAP and LacI (Fig.1). In the first mechanism (Fig.1a), both complexes are capable of interacting alternately with P\textit{lac}. The phenotypic outcome of this situation is dictated only by the time of presence of LacI, which is determined by the equilibrium constant of the LacI-LacO interaction. This mode of lac repression has been described through different approaches (Ozbudak et al., 2004; Garcia and Phillips, 2011) similarly assuming graded fractional occupation times, i.e. which can take any intermediate real value between 0 and 100\%. These graded mechanisms do not forbid a certain degree of heterogeneity in case of slow LacO-LacI interactions leading to transcriptional bursts which could be the rate-limiting molecular events for reaching a switching threshold (Choi et al., 2008). The limit case considered here is derived from an experiment of DNA immunoprecipitation presented below, showing that the association between LacI and P\textit{lac} can be very long-lived. In this extreme hypothesis, \textit{lac} induction is proposed to result from an event of complete LacI dissociation and to depend on the number of transcription initiation events taking place before reformation of a repression complex. Of course using the LacI-LacO equilibrium constant, as in (Ozbudak et al., 2004), would be inappropriate if a single dissociation event is sufficient to trigger the \textit{lac} switch, giving rise to a purely kinetic mechanism. When P\textit{lac} is cleared after dismantlement of a repressor complex, molecular races begin between LacI and RNAP to bind to DNA, possibly generating large transcriptional bursts (Choi et al., 2008). Alternatively, successive slow repression-derepression cycles can give bursts of proteins by translational amplification, even when a single mRNA is produced per derepression window (Choi et al., 2010). A simple possibility considered here is that a race occurs between one event of LacI rebinding and a certain number of RNAP-mediated transcription rounds. The outcome of this race determines if \textit{lac} is induced or not. In this scheme, two types of noninduced bacteria coexist at low inducer concentration (Fig.1b): those in which LacI did not dissociate from P\textit{lac} and those in which LacI dissociated from, but rebound to P\textit{lac} before the achievement of a threshold number of transcription events, thereby reinitializing the chain. Conversely, the induced bacteria are those which overcame two successive conditions: (\textbf{i}) the complete dissociation of LacI and (\textbf{ii}) the absence of premature re-repression.
\section{The single random event}
Following the proposal of (Choi et al., 2008), the limiting step for reaching the \textit{lac} expression threshold necessary for switching to the high expression state could be the complete dissociation of LacI. Although the release of LacI from P\textit{lac} is the keystone of \textit{lac} regulation, in fact little is known about the precise molecular mechanisms underlying this dissociation. The role initially attributed to the inducer was to force LacI to dissociate from DNA. Such an active dislodgement mechanism would however require some complex intramolecular energy transfer. A simpler scenario could be that LacI occasionally and spontaneously dissociates and is then prevented to rebind by the inducer, when present (Jacob and Monod, 1961; Choi et al., 2008). A third possibility exists. If the conformations of the different subunits of LacI are not mutually constrained, the repressor could fix the inducer while still bound to DNA thanks to its multimeric structure. Indeed, LacI is an homo-tetramer made of two linked homodimers, each one binding to a LacO palindromic site. The LacO palindrome is itself made of two half-sites, each one interacting with a LacI protomer (Wilson et al., 2007). The LacI tetramer can bind simultaneously to two operators in P\textit{lac}, with formation of a DNA loop possibly strengthened by DNA-bending proteins. The different operators of P\textit{lac} have been shown to cooperate for repression (Oehler et al., 1990) and to be involved in the genesis of the sigmoidal shape of the response to inducer (Oehler et al., 2006). These studies were however interpreted by using a continuous treatment that is not compatible with the discontinuous single event hypothesis. It is postulated here that the binding to either LacO or to the inducer are mutually exclusive for a single LacI protomer, but not for tetrameric LacI whose different protomers can have different binding states. Based on this hypothesis, the association between a LacI tetramer and P\textit{lac} can be very long-lived, as the tetramer is embedded into a macromolecular repression complex comprising a stabilized DNA loop, two \textit{lac} operators and accessory molecules. In vitro studies allowed to measure affinity constants between LacI and LacO, to determine that the DNA-binding unit of LacI is the dimer (Chen and Matthews, 1994) and that there are extensive monomer-monomer interactions (Lewis, 2005). However, a general problem in biochemical modeling is that kinetic and thermodynamic parameters determined in vitro are rarely transposable to in vivo conditions (Beard, 2011). This is particularly true for \textit{lac} because additional factors could stabilize the LacI-LacO loop in vivo, including DNA supercoiling and DNA-bending proteins, such as cAMP receptor (or catabolic regulatory) protein CRP (Lewis, 2005; Kuhlman et al., 2007), HU (Becker et al., 2007) or IHF. A loss of IHF has been observed upon \textit{lac} induction (Grainger et al., 2006). Negative supercoiling has been predicted to render the loop complex essentially nondissociable (Brenowitz et al., 1991). Finally, an overlooked aspect of CRP, beside its known role in stimulating \textit{lac} transcription after LacI release, is to repress \textit{lac} expression by facilitating LacI-mediated DNA looping (Fried and Hudson, 1996; Balaeff et al., 2004; Kuhlman, Zhang et al. 2007). This ambiguous action of CRP could explain the complexity of the combined effects of cAMP and \textit{lac} inducer (Setty et al., 2003; Kuhlman et al., 2007).\\
\begin{center}
\includegraphics[width=7.7cm]{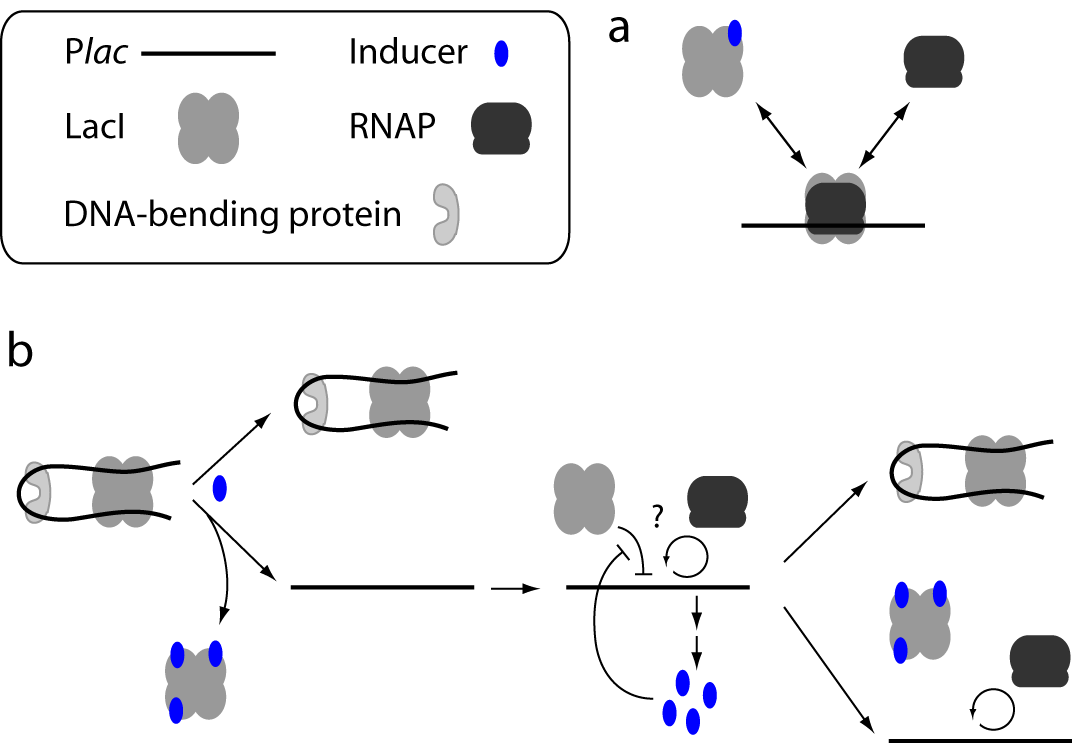} \\
\end{center}
\textbf{Figure 1}. \begin{small} Models of inhibition of \textit{lac} transcription through P\textit{lac} occupation by LacI. (\textbf{a}) RNAP can access DNA during the periods of absence of LacI, determined through a steady-state fractional occupation approach. LacI interaction with a single operator (O1) overlapping P\textit{lac}, is sufficient for this graded mode of repression. (\textbf{b}) In this alternative picture, the transcriptional competence of P\textit{lac} is determined by rare events of disruption and reformation of repression complexes. In this binary mode of repression, only stable repression complexes are taken into account, with two operators and a DNA hairpin stabilized by DNA-bending proteins.\end{small}\\
\newline
These collective molecular influences could make the apparent (re)-association rate of each LacI protomer, much higher than the dissociation rate, thus forbidding molecular escape. The probability of simultaneous dissociation of all the protomers roughly corresponds to the product of the probabilities of dissociation of the individual protomers, which can be extremely low. One step beyond, one cannot exclude that these stable interactions can be preserved during replication, in the same manner that in eukaryotes, certain nucleosomes split and duplicate during replication (Xu et al., 2010) and heterochromatin (repressive for gene expression) can be written out during mitosis (Moazed, 2011). The presence of 3 operators in the \textit{lac} operon, together with the multimeric nature of LacI, appear well suited to perpetuate the \textit{lac} repression complex during cell division. In this context, the \textit{lac} repressor complex would disassemble only upon inducer addition. To get an idea of the stability of the LacI-DNA interaction in vivo, the resistance of purified LacI-P\textit{lac} complexes formed in vivo was tested in vitro under stringent washing conditions. To this end, an indirect DNA immunoprecipitation method analogous to a chromatin immunoprecipitation assay (ChIP) was used. As shown in Fig.2, the molecular complex including the \textit{lac} operator and LacI can be isolated and preserved in absence of chemical crosslinking, a phenomenon which has never been reported so far for any transcription factor. This P\textit{lac}-LacI interaction pre-built in vivo, is shown to survive in vitro many cycles of washing and tube transfers, during 72 hours and in presence of 0.5 M NaCl (Fig.2). Such a stability is exceptional compared to classical DNA-protein interactions measured in vitro. This ChIP-like experiment strongly supports the possibility that in vitro and in vivo dissociations rates can be very different. \\
\end{multicols}
\begin{center}
\includegraphics[width=11cm]{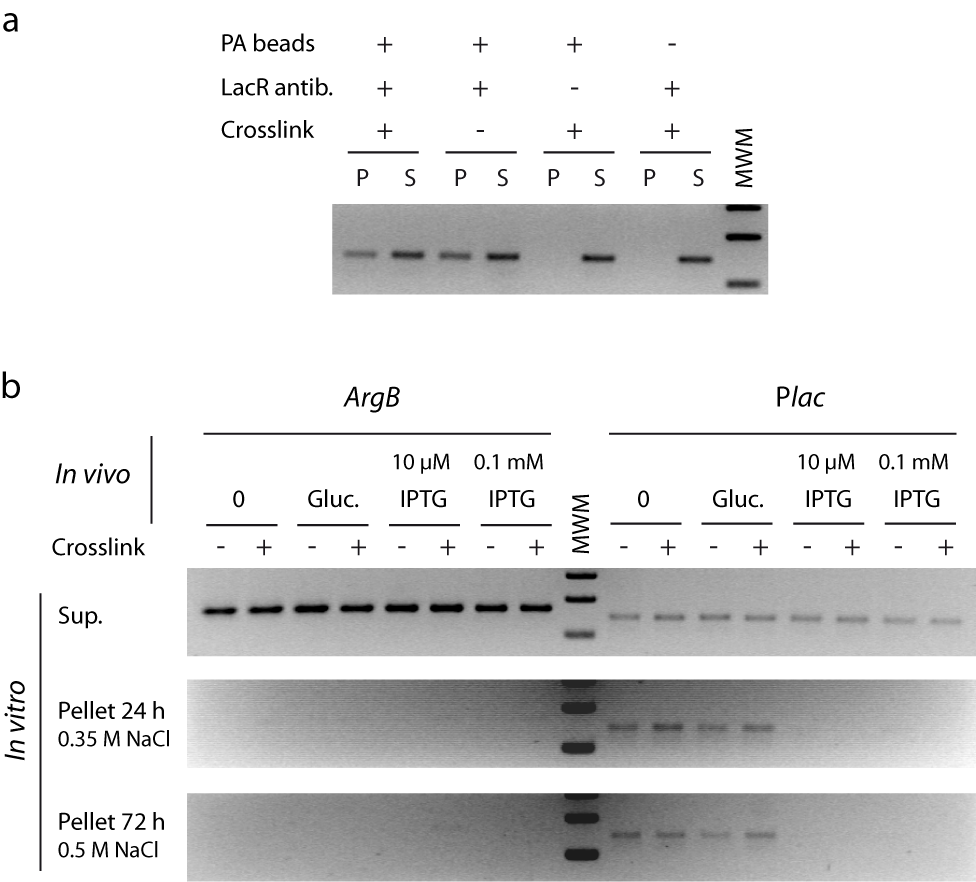} \\
\end{center}
\textbf{Figure 2}. \begin{small} Stability of the LacI-LacO interaction. LacI-DNA complexes were prepared from \textit{E. coli} K12 cells submitted or not to chemical crosslinking using formaldehyde, and the persistence of complexes was tested after two day-long washing. (\textbf{a}) P\textit{lac} DNA fragment was immunoprecipitated by incubation of LacI antibody with an excess of cellular extracts (P, pellet; S, supernatant). P\textit{lac} immunoprecipitation was no longer obtained in absence of protein A-coated beads (PA-beads), ruling out nonspecific adsorption to tube surfaces, and in absence of LacI antibody, ruling out nonspecific binding to the beads. (\textbf{b}) The persistence of LacI-Plac interaction was tested by high stringency washing of the DNA protein complexes attached to PA beads. A total of 4 tube transfers and 10 cycles of supernatant removal and replacement with fresh washing medium, were used for the final condition.\end{small}\\
\begin{multicols}{2}
\section{Sigmoidal behaviors of LacI dissociation can result from a chain of stochastic events}

In absence of precise knowledge about the organization of the LacI repressor complex, a very simplified hypothetical chain of events following inducer addition is proposed in Fig.3b, beginning with a spontaneous dissociation from one of the four LacO half-sites, likely to correspond to the less perfect one. This event would immediately reverse unless the dissociated protomer is sequestered by a molecule of inducer. Hence, the concentration of the inducer in the cell is critical for maintaining this essential step. The same process should repeat for the second and third inducer fixation, until the 7th step and the complete clearance of P\textit{lac}. If the 4 operator half-sites were strictly identical, this chain would have been disordered, but since this is not the case, a predominant ordered scheme will be considered, corresponding to the inverse ranking of half-site affinity for LacI. This stepwise mechanism is strongly supported by the suboptimal organization of the natural LacO sequence, compared to the artificial one used in (Sadler et al., 1983; Simons et al., 1984) and whose perfect symmetry was shown to enhance LacI binding. The defective spacing between half-sites is likely to create a tension between contiguous LacI protomers, forcing one of them to frequently dissociate. In this respect, it seems reasonable to suppose that the suboptimal organization of LacO would not have been evolutionary selected if the fixation of inducer to DNA-bound LacI was sufficient to remove it through an allosteric transition. In the scheme of Fig.3b, the single event postulated by (Novick and Weiner, 1957) is proposed to precisely correspond to the 7th step of the chain, instead of the synthesis of a single permease imagined by these authors. This step is a prerequisite for priming the switch but may not always trigger the switch, owing to an induction threshold (section 5.2.5). Let us first consider how this chain of events can explain the observed responses of bacterial cultures to inducer addition.

\subsection{Sigmoidicity in time}
\textit{Lac} induction is clearly sigmoidal in time, particularly at low inducer concentration, as illustrated by the Fig.2 of (Novick and Weiner, 1957) obtained at steady state bacterial concentration. Sigmoidicity in time can have several origins. The establishment of positive feedbacks is typically sigmoidal along time, irrespective of the existence of a threshold. In this case, sigmoidicity emerges from reiterative cycles of amplification whatever the number of relays involved in the circuit and even with hyperbolic production functions. But this explanation cannot be retained here since individual \textit{lac} switches are quasi-instantaneous (Vilar et al., 2003; Choi et al., 2008). Alternatively, as proposed here, sigmoidicity in time can also emerge from the averaged occurrence of a single event at the population level, in relation with the delay imposed by a chain of previous events. In the model presented below, the probability of complete release of LacI from LacO is conditioned by the sequence of events detailed in Fig.3b, where certain transitions depend on the concentration of the inducer ($ A $). If the triggering event was not causally subordinated to preliminary events or if a single transition in the chain was much more limiting kinetically than the other ones, then, the process would not have been sigmoidal. The probability of occurrence of this single molecular event is visible only at the population level. The positive feedback is not involved in the observed sigmoidicity of the populational \textit{lac} response, but only explains how a single event can trigger the switch. Before reaching the 7th state ($ x_7 $ in Fig.3b), the LacI/DNA complex should first pass through all the previous intermediate states. The total number of jumps can dramatically increase if backward transitions are more probable than forward ones, which is the case in absence or presence of low amounts of inducer in the cell, so that LacI dissociates very rarely. By contrast in presence of increasing concentrations of inducer, the probability of the 2d, 4th and 6th transitions can approach, equalize or exceed the reverse ones. \\

\subsection{Biochemical delay between inducer addition and the complete dissociation of LacI from P\textit{lac}}
In addition to the 15 transitions represented in Fig.3b, many elementary events are expected to take place between inducer addition and LacI disengagement (Appendix A1), which can be represented in the theoretical form of Fig.3a. The rates $ k $ are labeled with + or - depending on whether they correspond to forward or backward transitions, and the suffix numbers refer to the starting states. In this system, the mean time of arrival, holding for any chain length, is
 \\
\begin{equation} \left \langle T \right \rangle=\sum_{h=0}^{n-1}\sum_{i=0}^{n-h-1}\frac{1}{k^{-}_{i}}\prod_{j=i}^{h+i}\frac{k^{-}_{j}}{k^{+}_{j}} \end{equation}
 \\
which simplifies when all the backward rates are the same $  (k^{-})$ and the forward rates are the same $ (k^{+})$,
\begin{equation} \left \langle T \right \rangle=\sum_{i=0}^{n-1}(n-i)(k^{-})^{i}/(k^{+})^{i+1} \end{equation}
This value further simplifies if all the reaction rates are identical $ (k^{-} = k^{+}  = k) $,
\begin{equation}  \left \langle T \right \rangle= \frac{n(n+1)}{2k} \end{equation}

The mean time defined in Eq.(1) is relatively complicated but the corresponding probability distribution is even more awkward, as illustrated by the simple example of the two-step reaction shown in Appendix A2d.

\begin{center}
\includegraphics[width=8cm]{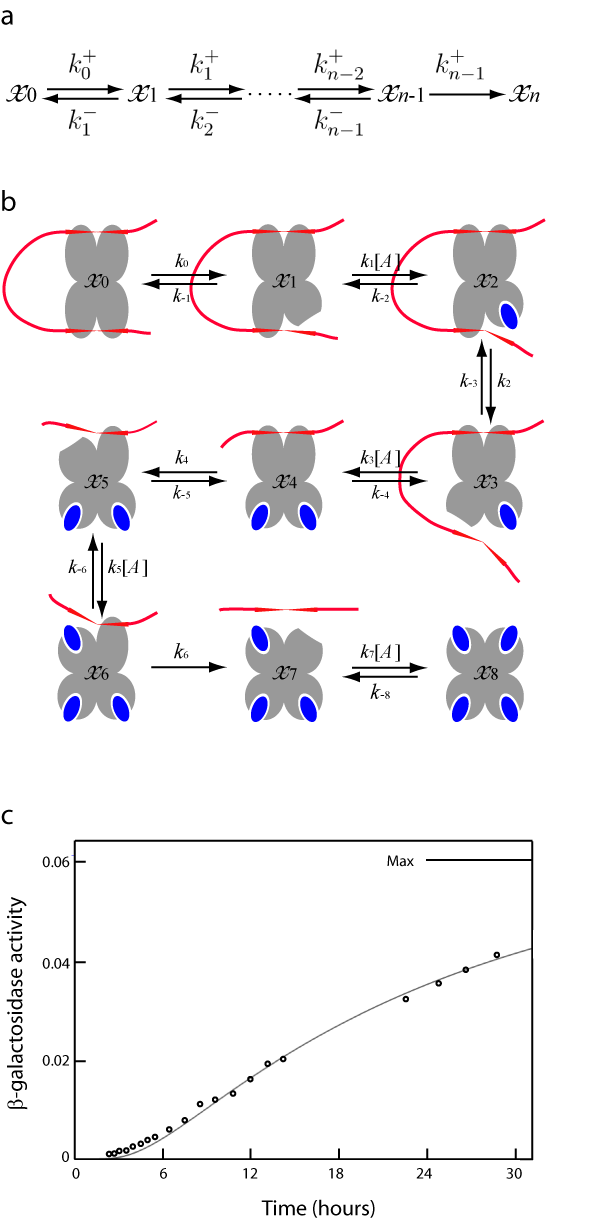} \\
\end{center}
\textbf{Figure 3}. \begin{small} Stochastic chain leading to LacI dissociation. (\textbf{a}) \textit{n} step-random walk. The initial state $ x_0 $ is a reflecting boundary while $ x_n $ is an absorbing state ($ k_n^{-} = 0 $). (\textbf{b}) Simplified scenario of dissociation of a LacI tetramer from two palindromic operators (opposite red arrows). In this stepwise mechanism, the random event triggering lac expression is the 7th step, which is itself subordinated to the achievement of the preceding chain of events. The 7th event can be considered as micro-irreversible in practice, considering the low concentration of LacI and the long waiting time before rebinding. The DNA binding steps $ k_1^- $ and $ k_3^- $ are approximated as first-order conformational transitions independent of the concentration of LacI in the cell. "\textit{A}" is the inducer (activator). The chain of successive LacO half-site clearances postulated in this scheme is only the predominant one, following the same order that the LacI affinity ranking, so that no statistical balancing of the microscopic rate constants is used for simplicity. (\textbf{c}) Accumulation of induced bacteria (circles) at low concentrations of inducer (Data from Novick and Weiner, 1957). Plain line drawn to the probability of completion of a large random walk (Eq.(4)). \end{small}\\

To simplify the question, let us assume that in the course of the switch and for a critical concentration of inducer, there could be a particular situation in which the resistance and the tendency to dissociation roughly equalize and the process becomes a random walk. The highest entropy random walk, called isotropic or symmetric, is obtained when all the transitions are equally probable, so that the degree of uncertainty about the position of a walker is maximal. Using this approximation and according to the central limit theorem, the probability of occupation of the different states follows a Gaussian distribution. For long chains, all mathematical methods converge towards a similar expression of the probability to reach the state $ x_n $, as function of the dimensionless intrinsic time of the walk $\theta$.

\begin{subequations} \label{E:gp}
\begin{equation} P_{n}(\theta )=1-\frac{4}{\pi}\sum_{j(odd)=1}^{\infty }\frac{(-1)^{\frac{j-1}{2}}}{j}\e^{-\frac{\pi ^2j^2}{8}\theta  } \end{equation} \label{E:gp1}

where $\theta$ can be defined in multiple ways depending on the approach. In the discrete random walk approach,
\begin{equation} \theta = N/n^2 \end{equation} \label{E:gp2}
In the time continuous approach,
\begin{equation} \theta = 2kt/n^2 \end{equation} \label{E:gp3}
or
\begin{equation} \theta = t/\left \langle T \right \rangle \end{equation} \label{E:gp4}
\end{subequations} 
(Appendix A1). Whatever the method used, Eq.(4) is a sigmoidal function of either \textit{N} or \textit{t}. The elimination of the number of elementary steps could make Eqs.4a/4d, a satisfying universal substitute for approximating the evolution of long isotropic chains whose explicit calculation is impossible. If the rates are not too different from each others, and if assuming that LacI-LacO interactions are preserved during DNA replication, fitting this theoretical probability distribution to the experimental points of Novick and Weiner leads to a mean time of about 27 hours for 7 $ \mu $M of the inducer methyl-1-thio-$ \beta $-D-galactopyranoside (TMG). However, this fitting is approximate since the isotropic random walk is an oversimplified hypothesis and the concentration of inducer could not correspond precisely to the point of equivalence between forward and reverse rates. In addition, all the bacteria in which LacI dissociates from P\textit{lac} do not necessarily become induced (section 5.2.5), so that the Fig.2 of Novick and Wiener could superpose bacteria whose successful windows of derepression are spread out in time. Nevertheless, the matching between these curves and these points (Fig.3c) is satisfying enough to make plausible this origin of sigmoidicity.

\subsection{Increasing the response tightness at high inducer concentration}

Increasing the inducer concentration strongly reduces the delay of response and increases the tightness of its time distribution. This latter effect could be related to the fact that pseudo-first order rates largely exceed the rates of the corresponding backward transitions (Fig.3b), and could reflect the well-established focusing effect of stepwise oriented mechanisms. Subdividing an event of given duration into many hierarchical sub-events whose sum has the same duration, increases the complexity of the global probability density, but also the reproducibility of the total completion time. Indeed, the variance of a multistep chain made of consecutive events, exponentially spaced with variance $ 1/k^2 $, is lower than that of a single stochastic transition of identical mean waiting time since
\begin{equation} \sum_{j=1}^{n}\left (\frac{1}{k_{j}} \right )^2< \left (\sum_{j=1}^{n}\frac{1}{k_{j}}  \right )^2 \end{equation}
The focusing effect of multiple stochastic events has been invoked to explain the tightness of reaction time of a stepwise mechanism such as the reproducible photoreceptor responses to single-photons arising from a series of consecutive Poisson events (Doan et al., 2006).

\subsection{Inducer concentration-dependent sigmoidicity }
\textit{Lac} induction is a sigmoidal function of inducer concentration, as experimentally observed both in vivo for bacterial cultures (Fig.4a) and in vitro when LacI is bound simultaneously to two operators (Oehler et al., 2006). 
\begin{center}
\includegraphics[width=7.5cm]{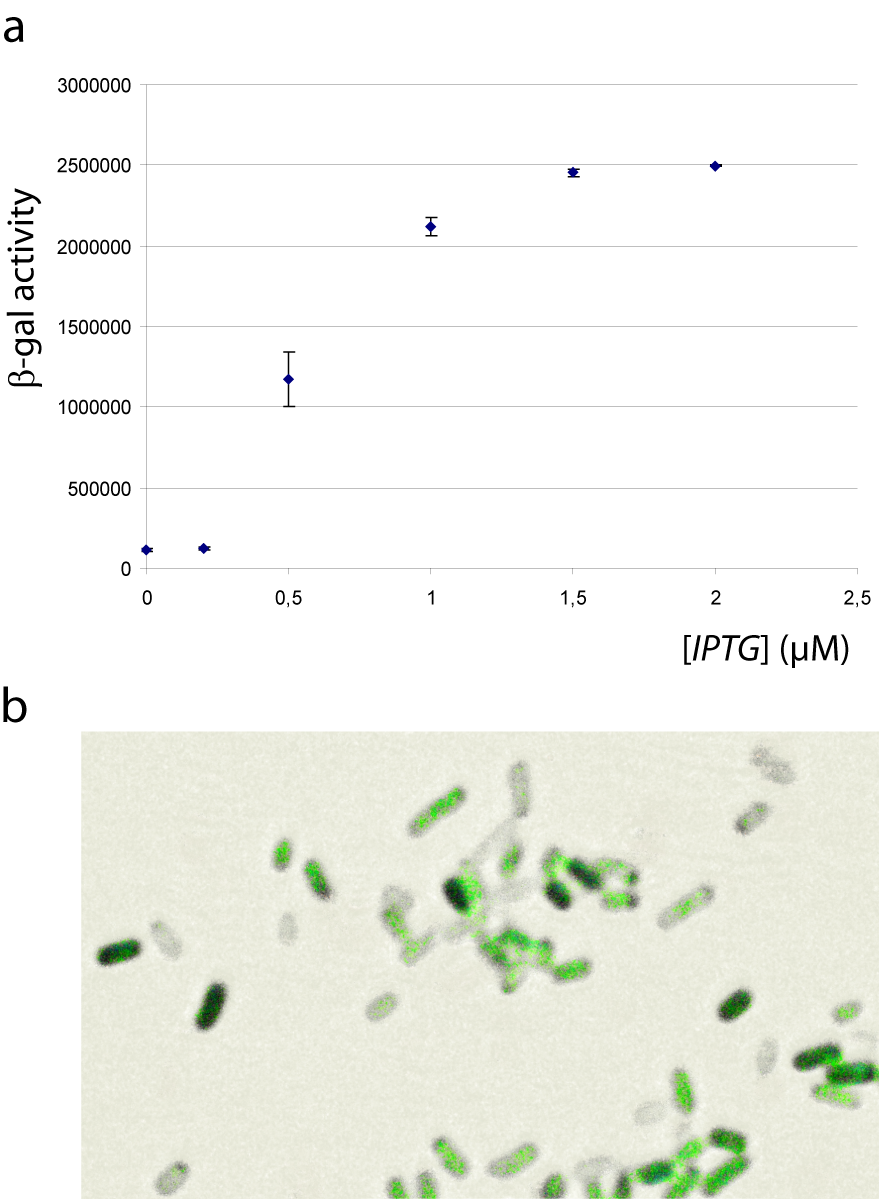} \\
\end{center}
\textbf{Figure 4}.\begin{small} (\textbf{a}) $ \beta $-galactosidase activity in bulk assay is a sigmoidal function of inducer. (\textbf{b}) Heterogeneous $ \beta $-galactosidase staining obtained at the single cell level with 0.5 $ \mu $M IPTG. The intermediate staining could be due to the basal expression rate $ k_0 $ postulated in presence of LacI and, after LacI release, to the progressive accumulation of $ \beta $-galactosidase before reaching the steady state value $ r/(k_0 + k_1) $. \end{small}\\
\newline
Concentration-dependent sigmoidicity is the classical nonlinearity condition retained for generating bistability from a positive circuit (Chung and Stephanopoulos, 1996; Cherry and Adler, 2000); but no clear consensus emerged about the existence of such a cooperativity for \textit{lac}.

\subsubsection{Quasi-equilibrium sigmoidal function of inducer concentration}
Ligand-dependent sigmoidicity is generally thought to result from nonindependent multiple binding (O'Gorman et al., 1980; Sharp, 2011). Collective binding processes are numerous in the \textit{lac} system. \textit{Lac} repressor is a tetramer, which can bind to several inducer molecules (Ozbudak et al., 2004; Kuhlman et al., 2007), and to different operators in P\textit{lac} (Oehler et al., 1990; Oehler et al., 2006; Daber et al., 2009). Hill and cooperative Adair saturation functions are generally selected to define production functions (Chung and Stephanopoulos, 1996; Ozbudak et al., 2004; Yildirim and Kazanci, 2011). Of course, the sets of differential equations (ODEs) with such production functions predict \textit{lac} bistability, but this modeling is questionable from the single event hypothesis and considering the poor binding cooperativity reported for the \textit{lac} system components. In this respect, analyzing any sigmoidal curve with "observed Hill coefficients", is not really relevant when sigmoidicity does not relate to a genuine Hill process. An alternative mechanism is proposed below.

\subsubsection{Concentration-dependent transient sigmoidal responses}
The stepwise process described in Fig.3b starting from the hairpin repression complex, allows transient sigmoidicity to arise from the combination of pseudo-first-order functions of the inducer concentration. For a given time after inducer addition, the probability that LacI dissociates from a LacO hairpin complex increases in a sigmoidal function of inducer concentration. The simple example of a two-step reaction shown in Appendix A2d, is sufficient to display sigmoidicity. Deriving phenomenological Hill coefficients from such curves would be obviously nonrelevant. In absence of DNA loop, the induction of \textit{lac} in vivo, as well as the clearance of P\textit{lac} in vitro, were shown to take a nonsigmoidal shape (Oehler et al., 2006). This finding has been interpreted through the existence of allosteric cooperativity, but it could result as well from the modification of the number of steps in the dissociation chain. This kinetic mode of ultrasensitivity to inducer concentration is not comparable to the cooperative one invoked to generate bistability.

\section{LacI dissociation envisioned as a turning point in the \textit{lac} system}

The event of disruption of the LacI-mediated repression complex is embedded in the general \textit{lac} circuit detailed below.

\subsection{The \textit{lac} circuit}
Once LacI is released from DNA, RNAP can repeatedly access P\textit{lac} and initiate successive rounds of \textit{lac} mRNA transcription through a hit-and-run mechanism, at approximately constant rate $ k_1 $, considering the relative invariance of RNAP concentration in the cell. The value of $ k_1 $ corresponds to that of the pseudo-first-order rate of productive collision between RNAP holoenzyme (including the initiation factor $ \sigma 70 $) and P\textit{lac}, followed by a rapid and driven elongation process. The evolution of \textit{lac} mRNA results from a chain of connected events detailed below, in which the parameters subject to transient evolution immediately after LacI release will be labeled with (\textit{t}). \textit{Lac} mRNAs (\textit{M}) evolve following Eq.(6), where $ k_0 $ is the basal transcription rate in presence of LacI and CRP, and $ r_M $ is the rate of removal of \textit{lac} mRNAs. 
\newline
\begin{equation} \frac{dM(t)}{dt}=k_{0}+k_{1}P(t)-r_{M}M(t) \end{equation}
\newline
In GRN modeling, the maximal transcription rate is weighted by a probability, generally that an activator is present, or that a repressor is absent (Michel, 2010). In the case of \textit{lac}, this probability written \textit{P}(\textit{t}) in Eq.(6), can have different meanings depending on the model used and will be defined through several functions $(f_1)$. For comparison, \textit{P}(\textit{t}) will be defined as either (\textbf{i}) the steady state fractional time of absence of the repressor (involving many events) under the classical graded repression hypothesis, or (\textbf{ii}) the probability that LacI does not rebind prematurely under the single event hypothesis (5.2.5). In all cases, this probability is a function of \textit{lac} expression, leading to permease synthesis, allowing the entry of more inducer in the cell, that ultimately decreases the probability of LacI rebinding, in a circular relationship. \textit{Lac} proteins ($ L $) follow
\newline
\begin{equation} \frac{dL(t)}{dt}=s\ M(t)-r_{L}L(t) \end{equation}
\newline
where \textit{s} and $ r_L $ are the rates of synthesis and removal of the products respectively. For simplicity, all \textit{lac} proteins, including permease and $ \beta $-galactosidase, are supposed to be degraded at the same rate. The intracellular amount of inducer written $ A $ (activator), depends on the passive, reversible and saturable transport of external inducer through permease, classically described by
\newline
\begin{equation} \frac{dA(t)}{dt}=vL(t)\left (\frac{A_{out}}{K_{o}+A_{out}}-\frac{A(t)}{K_{i}+A(t)} \right ) \end{equation}
\newline
The concentration of intracellular inducer then directly regulates the fraction of total repressor that is competent for binding DNA ($ R_{c} $), through a function $ f_2 $
\newline
\begin{equation} R_{c}(t)=f_{2}\left (R_{T},A(t)   \right ) \end{equation}
\newline
for which, once again, several definitions will be envisioned depending on the model used and $ R_T $ is the total concentration of LacI, supposed to be roughly constant. Simplifying hypotheses can be used to define this function. In particular, inducer concentration can be considered as higher than the low concentration of LacI in the cell, so that it can be incorporated into pseudo-first-order binding rates and the amounts of free and bound inducer can be approximated as equivalent. It is important to note that the probability that LacI does not rebind also depends on time and on \textit{lac} products, including (\textbf{i}) permease which regulates the intracellular concentration of inducer and (\textbf{ii}) $ \beta $-galactosidase for metabolizable inducers. Since it is supposed to fluctuate slowly relatively to gene expression, $ R_{c} $ will be incorporated into the pseudo-first-order rate of LacI binding to LacO, such that $ k_{2}(t)=k^{*}_{2}R_{c}(t) $, where $ k_2^* $ is a second-order binding rate. The low number of repressors in the cell led certain authors to distinguish between their free and total concentrations (Yagil and Yagil, 1971; Berg and Blomberg, 1977), but considering that there is a unique combination of  strong operators in the cell, the total number of LacI ($ VR_{c} $) and the number of LacI not bound to LacO ($ VR_{c}-1 $) will be considered as equivalent, irrespective of whether LacI jumps in solution or slides along DNA. A refinement in this system is that contrary to $ k_1 $, $ k_2 $ is not constant but decreases with time since $ R_{c} $ depends on the concentration of inducer, that is itself a function of the number of permeases in the cell. To evaluate the general behavior of the system, further simplifications can be made. The \textit{lac} products (including mRNA and proteins) are fused together and collectively written $ x(t) $, by assuming that transcription elongation and translation of \textit{lac} mRNAs (occurring simultaneously in bacteria) are quasi-instantaneous. In this respect, note that population heterogeneity has also been visualized at the level of \textit{lac} mRNA (Tolker-Nielsen et al., 1998), suggesting that it is not primarily due to translational bursts. The import of inducer by permease is supposed to be far from saturation and its export is considered as negligible soon after LacI release. To complete the system, it is now necessary to define the functions $ f_1 $ and $ f_2 $ postulated above to describe $ P(t) $ and $ R_{c}(t) $.

\subsection{The question of \textit{lac} bistability}
Multistability is a general mechanism to explain how nonunique cell types can stably coexist within the same environment, but the origin, and even the existence of \textit{lac} bistability, raised active debates and remain open questions (D\'iaz-Hern\'andez and Santill\'an, 2010; Veliz-Cuba and Stigler, 2011; Savageau, 2011). The notion of bistability is delicate in this system since the bimodal distribution of bacteria, induced and uninduced, is observable only for a particular, low range of inducer concentration and it is difficult to certify experimentally that this bimodality is really stationary rather than extended transient. If the induced bacteria seem indeed to remain in their state for generations, it is less clear that uninduced bacteria cannot switch to the induced state. In addition, using a low inducer concentration, as for the Fig.2 of Novick and Wiener reproduced here in Fig.3c, bacteria should be examined very long after inducer addition and renewal to consider that a steady state is reached.

\subsubsection{The logic of \textit{lac} bistability}
If postulating that LacI frequently dissociates from LacO according to a dynamic model, a bistability threshold is required since otherwise the feedback would be primed by low stochastic \textit{lac} expression as soon as the inducer in present. A "priming threshold" of \textit{lac} expression is necessary to ensure the relative stability of the uninduced population fraction. Such a threshold before ignition is typically provided by multistability (Chung and Stephanopoulos, 1996) and is all the more necessary for \textit{lac} that a basal expression is a prerequisite for \textit{lac} induction. In fact, \textit{lac} expression is an all-or-almost-nothing phenomenon. In the population assay shown on Fig.4a, a low but reproducible basal \textit{lac} activity, of about 4\% of the maximal activity, is observed. The homogeneous pattern of LacZ staining suggests that this expression cannot be attributed to a fraction of induced bacteria, but to a fraction of maximal \textit{lac} expression in every bacterium, whose rate is written $ k_0 $ in Eq.(6). This basal expression seems at first glance contradictory with the all-or-none hypothesis but is in fact necessary for allowing the bacteria to sense the availability of lactose in their environment. This low level of transcription in presence of LacI could for instance correspond to the small bursts described in (Choi et al., 2008), or to cAMP-liganded CRP-promoted transcription (Kuhlman et al., 2007). This basal activity is both compatible with and necessary for \textit{lac} induction. Indeed, (\textbf{i}) Basal \textit{lac} expression is necessary to generate small amounts of permease/LacY for allowing the entry of the first molecules of inducer. (\textbf{ii}) In natural conditions, basal \textit{lac} expression is also necessary to generate small amounts of LacZ, to convert these first molecules of lactose into allolactose, which is the genuine \textit{lac} inducer (Wilson et al., 2007). The long-standing apparent paradox that \textit{lac} should be active to be activatable, is typical of bacterial regulatory networks and is also encountered for example in the case of the SOS regulon. But to be compatible with the model, basal \textit{lac} expression should be low enough to not prime the positive feedback circuit in all cells once inducer is present. This latter condition is classically fulfilled if the system is robustly bistable.

\subsubsection{The classical quasi-equilibrium model of bistability}
The ultrasensitivity of \textit{lac} expression to inducer concentration, with a sigmoidal or S-shaped input function, has been shown crucial for the all-or-none nature of the switch. \textit{Lac} expression is generally supposed to be governed by graded interaction cycles between LacI and LacO, that can be up- and down-regulated in a continuous manner, thus allowing to fine-tune the fraction of time during which LacO is unoccupied (D\'iaz-Hern\'andez and Santill\'an, 2010; Yildirim and Kazanci, 2011). To adapt the model of graded repression to \textit{lac}, one should distinguish two modes of interaction between LacI and P\textit{lac}: (\textbf{i}) the one previously described, holding before the initial dissociation, very stable and requiring two operators (section 5.2.5) and (\textbf{ii}) an other one, examined below (sections 5.2.3 and 5.2.4), reversible and more dynamic, between LacI and a single operator (O1), as schematized in Fig.1a. To illustrate the principle of graded repression, let us consider the typical example of bistability model described in (Ozbudak et al., 2004), one of the most frequently cited articles on the \textit{lac} system. The \textit{lac} products evolve according to
\newline
\begin{equation}\frac{dx}{dt}=k_{s}P-r_{x}x \end{equation}
\newline
where $ k_s $ is the maximal rate of synthesis, and $ r_x $ the rate of removal. $ P $ is the probability that LacI is not bound to LacO, depending on $ R_{c} $. The occupation of LacI by the inducer (\textit{A}) is rapid enough to be approximated as a quasi-equilibrium.
\newline
\begin{equation}P=\frac{1}{1+K_R R_{c}} \end{equation}
\newline
Eq.(11), where $ K_{R} $ is an equilibrium binding constant, describes a situation of graded repression in which $ P $ can take any real intermediate value between 0 and 1, even if slower interactions and transcriptional bursting could introduce some heterogeneity (Golding et al., 2005). $ R_{c} $ should now be determined. Its definition can be envisioned in two manners, with or without Hill cooperativity. 

\subsubsection{Quasi-equilibrium hypothesis with strong (Hill) equilibrium cooperativity}
Assuming that the concentration of \textit{A} largely exceeds that of LacI, a highly cooperative fixation of the inducer to LacI, postulated in (Ozbudak et al., 2004), is described by
\begin{equation}\frac{R_{c}}{R_{T}}=\frac{1}{1+(K_A A)^n}\end{equation}
Eq.(12) is a sigmoidal Hill repression function of the inducer concentration, reflecting strong cooperativity since it presupposes that LacI exists essentially in two forms, either devoid of, or filled with inducer. The decrease or the elimination of intermediate microstates characterise all cooperative phenomena (Michel, 2011). The approach to bistability using Hill functions long proved very successful (Cherry and Adler, 2000; Sobie, 2011) and it works well for \textit{lac}. Note that in addition to the Hill function, MWC equations were also used to describe allosteric binding (Appendix 2) and are also adequate for generating bistability.

\subsubsection{Quasi-equilibrium model without Hill cooperativity}
Given that the interaction between LacI and the inducer proved to be not or weakly cooperative, the different partially liganded microstates of LacI (Fig.5a) cannot be ignored and the ratio $ R_{c}/R_{T} $ should be determined without recourse to a Hill function. The general function describing nonallosteric interactions can be obtained by enumerating the microstates capable of binding to the bipartite operator O1. This palindromic sequence can be occupied in a graded manner by all LacI tetramers with two adjacent protomers devoid of inducer. They correspond to LacI microstates occupied by either 0, 1 or 2, but not 3 or 4 molecules of inducer. Moreover, only certain forms of di-liganded LacI can bind the operator, since at least one dimer in the tetramer should be devoid of inducer (group $ R_{c1} $ in Fig.5a). In this respect, fluorescent protein-tagged LacI, known to form dimers but not tetramers, have been shown sufficient to bind to LacO with short residence times (Elf et al. 2007).
\begin{center}
\includegraphics[width=7cm]{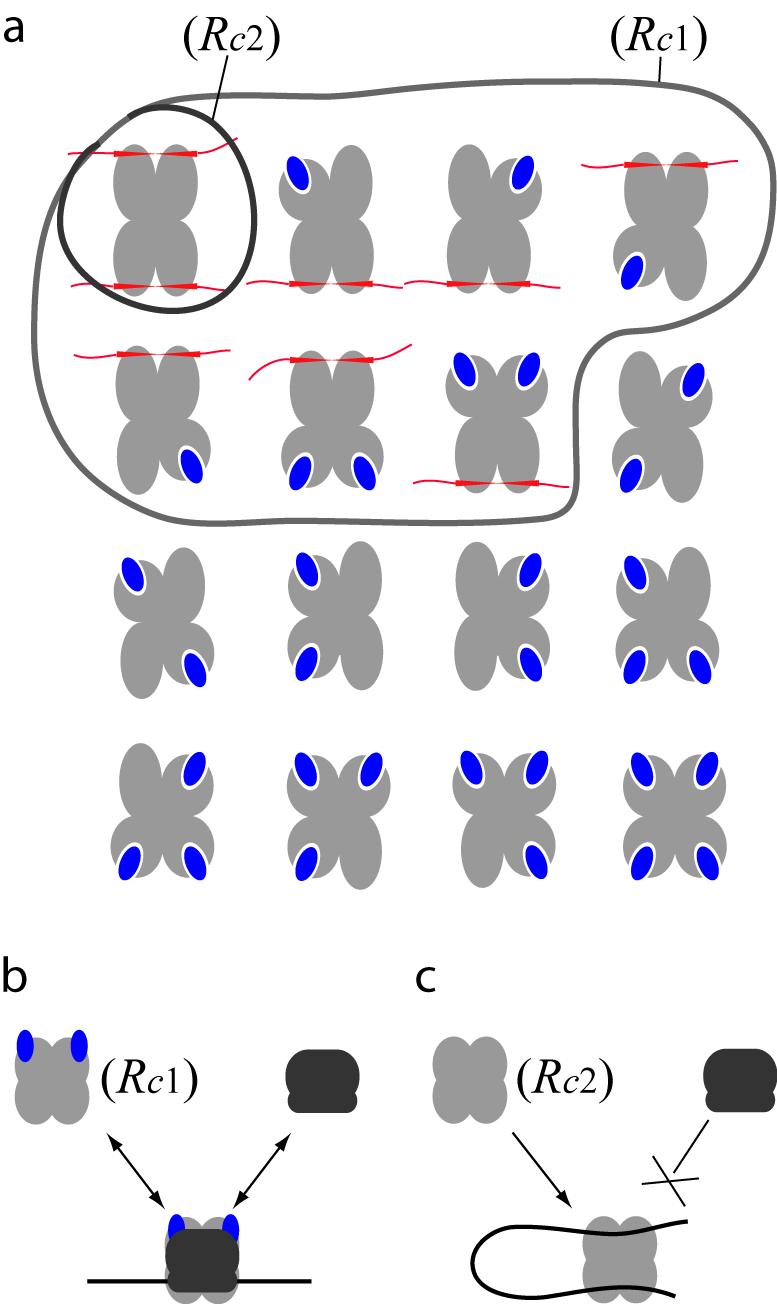} \\
\end{center}
\textbf{Figure 5}. \begin{small} (\textbf{a}) LacI tetramer microstates with respect to inducer binding. (\textbf{b}) 7 microstates ($ R_{c1} $) out of 16, are capable of interacting dynamically with a single palindromic operator, by sharing time with RNAP as supposed in classical \textit{lac} models. (\textbf{c}) Only one microstate ($ R_{c2} $) can form a stable hairpin through bridging two operators, thereby relocking P\textit{lac} and preventing RNAP to reinitiate transcription.\end{small}\\
\newline
When using the same microscopic binding constant ($ K_A $) for inducer binding to the different LacI protomers, the fraction of LacI defined above corresponds to 7 microstates out of 16 (Fig.5a). An additional subtlety in this enumeration, is that the binding rate of the microstate $ R_{c2} $ (Fig.5a, top left) should be multiplied by 2 because it can bind O1 in 2 different manners. If $ R_{T} $ is the total concentration of diffusing \textit{R} in the cell, supposed to be roughly constant, then $ R_{c1} $ can be written using an Adair formula as follows
\newline
\begin{equation} \frac{R_{c1}}{R_{T}}= \frac{2+4K_{A}A+2(K_{A}A)^2 }{(1+K_{A}A)^4}=\frac{2}{(1+K_{A}A)^2} \end{equation}
\newline
which could have been anticipated from the bilateral symmetry of the LacI tetramer. Bistability is no longer recovered using this value, suggesting that strong Hill-like cooperativity is required for the quasi-equilibrium mode of bistability. The graded mechanism described in sections 5.2.2 and 5.2.3 is very predominant in biochemical models but is not suited to the single event view and would be poorly compatible with the stability of the LacI-LacO interaction. Intriguingly, among the different mechanisms of equilibrium cooperativity available, the one sometimes retained in the context of \textit{lac}, is the popular Monod-Wyman-Changeux (MWC) model (O'Gorman et al., 1980; Sharp, 2011; Dunaway et al., 1980). This choice appears particularly not tenable under the hypothesis of delayed LacI rebinding described by Choi et al. (Choi et al., 2008), since it would lead to the complete release of LacI tetramers at every symmetric conformational change of the tetramer. A nearest neighbor sequential allosteric mechanism such as the KNF model (Koshland et al., 1966), would be more compatible with the quasi-irreversible dissociation model since it would explain how the departure of an inducer molecule from a LacI protomer, favors the dissociation from DNA of a neighboring partner. An alternative threshold mechanism can be proposed, that is relieved from the quasi-equilibrium assumption and the use of equilibrium constants.\\

\subsubsection{Modeling a single molecular race}
The traditional mode of bistability obtained in the previous sections, using ODEs with sigmoidal production functions containing equilibrium constants, is not expected to hold for P\textit{lac} in case of rare turnovers of the repression complex. Hence, an alternative model, purely kinetic, is proposed to give similar results for bacteria in which LacI dissociated from LacO. Based on the ChIP experiment, if a window of repression is long enough compared to the stability of \textit{lac} products, it could erase the results of the previous window of derepression which failed to trigger the induction and eliminate the inducer from the cell. Under this hypothesis, the capacity of bacteria to become induced or not should be determined during a single waiting time before reformation of this complex. This alternative mechanism is related to the question raised in (Choi et al., 2008): "Does every complete dissociation event lead to a phenotype transition?". In this situation, the system is not determined by a steady state, but by an initial rate, as postulated by Bolouri and Davidson for transcriptional cascades during development (Bolouri and Davidson, 2003) and whose approach will be partially picked up here. Contrary to classical GRN modeling in which the probability associated to the maximal transcription rate is defined by a steady state occupation ratio taking into account both the association and dissociation rates of a transcription factor, it is now the probability that LacI does not shut off again P\textit{lac} too soon, to allow a certain threshold number of transcription rounds to occur. The number of \textit{lac} expression events necessary to reach the point of no return of the switch will be called \textit{n}. Let be $ k_2 $ the pseudo-first-order rate (including the total concentrations of LacI supposed to be invariant), of P\textit{lac} relocking upon LacI-mediated re-building of an hairpin repression complex (section 3). Under the single event hypothesis, the positive feedback is a race between LacI and RNAP. To make the race fairer and to counterbalance their different kinetics due to the higher concentration of RNAP in the cell, $ n $ \textit{lac} expression rounds (total duration $ n/k_1 $) are supposed to be necessary during the mean time window of repression complex reformation ($1/ k_2 $). With given rate constants $ k_1 $ and $ k_2 $, this probability decreases when $ n $ increases (Fig.6a). The probability of $ n $ transcription events without any LacI relocking should be determined. At every time point, the probability that \textit{lac} is expressed is simply $ k_1/(k_1+k_2) $. Given the memoryless nature of this stochastic molecular race, the probability that it repeats $ n $ times consecutively is $ (k_1/(k_1+k_2))^n $. This intuitive result can be recovered rigorously from transient state probabilities as follows. The probability of no LacI relocking, which decreases exponentially with time with a density function $ k_2 \exp(-k_2 t)$, should be weighted by the probability of reaching the \textit{n}th expression round. This latter probability, written below $ P(X_1 \geq n) $, is the complement to one of the sum of probabilities of 0, 1, 2, .. until ($ n-1 $) hits,
\begin{equation} P(X_{1}\geq  n)=1-\sum_{j=0}^{n-1}P(X_{1}=j) \end{equation}
where $ P(X_1 =j) $ is a Poisson distribution of parameter $ k_1t $. Hence, the final probability of success is 
\begin{equation} \int_{t=0}^{\infty}k_{2}\e^{-k_{2}t}\left (1-\sum_{j=0}^{n-1}\frac{(k_{1}t)^{j}}{j!} \e^{-k_{1}t} \right )dt=\left (\frac{k_{1}}{k_{1}+k_{2}}  \right )^n \end{equation}
This probability corresponds to the proportion of cells that reach the induced state, in the cellular subpopulation in which LacI unhooked from P\textit{lac}. The direct calculation of \textit{n} is complicated by the fact, not taken into account above, that $ k_2 $ depends on \textit{lac} expression. Considering that the essential form of LacI capable of relocking the system is the microstate $ R_{c2} $ devoid of inducer (Fig.5a), the set of relationships described above can be compressed into the following deterministic equation
\begin{equation} \frac{dx}{dt}=k_{0}+k_{1}\left (\frac{k_{1}}{k_{1}+k_{2}\frac{1}{(1+\beta x)^4}}\right )^n-rx \end{equation}
with the initial condition, subject to stochastic variance, $ x_0 = k_0/r $ at $ t_0 $. For appropriate parameter ranges, this retroaction is typically capable of splitting the population into induced and uninduced cells (Fig.6b). Cooperative binding of inducer can still be introduced through a Hill function, simply by replacing $ (1+\beta x)^4 $ by $ 1+(\beta x)^4 $ in Eq.(16); but this postulate is not necessary to obtain the two populations, contrary to the previous quasi-equilibrium approach. This feature could be a sign that the kinetic approach is more realistic than the classical one in absence of obvious equilibrium cooperativity in the system. The negative integral of Eq.(16) resembles a sort of Waddington's landscape (Ferrell, 2012), in which the stochastic initial spreading of bacteria converges towards the two point attractors for \textit{lac} expression corresponding to the induced and uninduced states. Several points should be noted: (\textbf{i}) the bistable-like profile of Fig.6b is obtained because the system is restricted to a single LacI rebinding interval, and (\textbf{ii}) the evolution of \textit{lac} expression described by Eq.(16), is purely probabilistic and meaningless at the level of a single cell and is envisioned as the global behavior of a container of bacteria. In a given bacterium, the race is won by either LacI or RNAP. \textit{Lac} expression is binary in single cells and can take only two values (either $ k_0 $ or $ k_0+k_1 $), while the continuous Eq.(16) describes the graded variation of the fraction of inducible bacteria in the container. In addition, the splitting of the population into uninduced and induced states (Fig.6b) does not occur simultaneously for all the bacteria, since this process begins with the single event that is itself distributed along time, as described in section 4.2. In this development, the threshold $ n $ is not a predetermined constant to be reached, but is a commitment point regulated in concert with the combination of kinetic parameters of interaction.
\begin{equation} n=\frac{\ln (\rho x_{s}-\gamma )}{4\ln (1+\beta x_{s})-\ln (\alpha +(1+\beta x_{s})^4)} \end{equation}
where $ \gamma = k_0/k_1, \alpha = k_2/k_1, \rho = r/k_1 $ and $ x_s $ is the threshold amount of \textit{lac} products over which \textit{lac} is induced.
\begin{center}
\includegraphics[width=7.8cm]{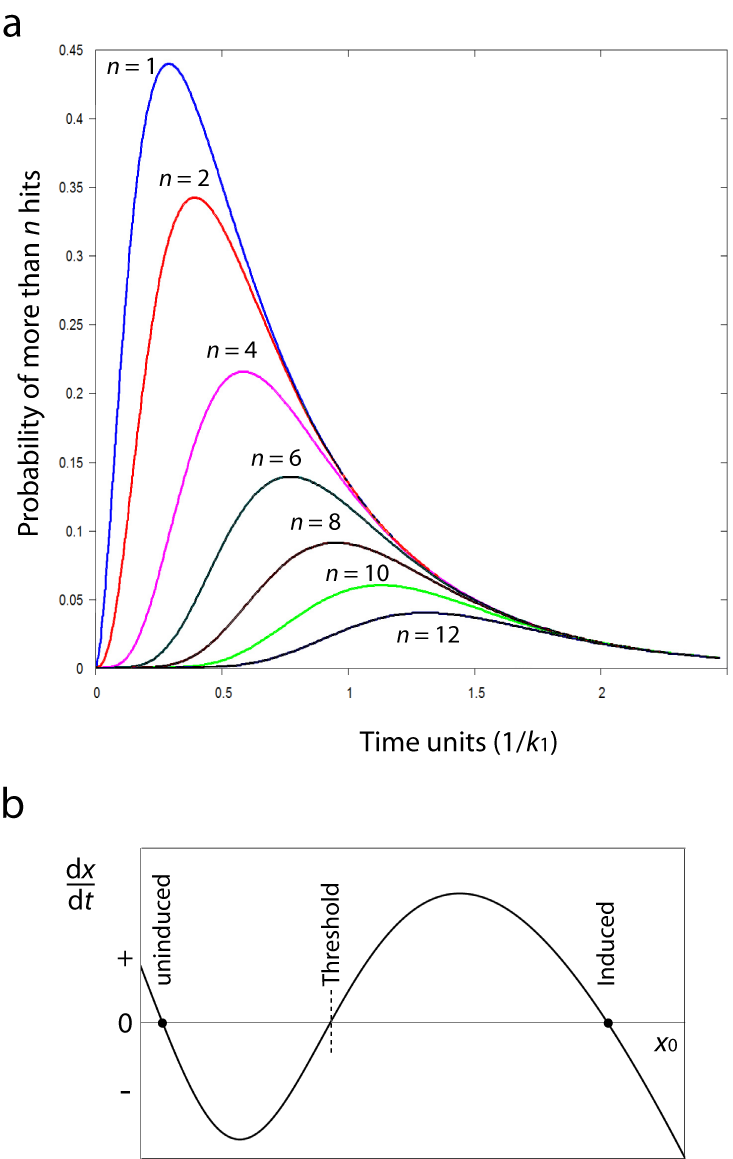} \\
\end{center}
\textbf{Figure 6}. \begin{small}(\textbf{a}) Probability that \textit{n} events occur before a single competing event (5 times slower in this example) (\textbf{b}) Probabilistic evolution rate of \textit{lac} products depending on their initial amounts, for a given inducer concentration (using Eq.(16) and the relative values, setting $ k_1=1, k_0=0.04, k_2=0.1, r = 0.2, \beta $= 0.4 and $ n $ = 100). Dots indicate the expected populations of bacteria, either uninduced or induced, after a derepression window. \end{small}\\
\section{Experimental methods}
\subsection{LacI-mediated P\textit{lac} DNA immunoprecipitation without crosslinking}
The stability of LacI-P\textit{lac} association was evaluated through a method equivalent to chromatin immunoprecipitation (ChIP) and performed using non-engineered \textit{E. coli} K12 cells (P4X strain), but with and without chemical crosslinking. The longevity of LacI-P\textit{lac} association was compared between (\textbf{i}) basal conditions, (\textbf{ii}) in long term presence of high inducer concentration (Isopropyl-$ \beta $-D-1-thiogalactopyranoside IPTG), expected to dismantle the repressor complex, and (\textbf{iii}) in presence of glucose (20 mM), possibly influencing LacI binding indirectly through CRP (Fried and Hudson, 1996; Balaeff et al., 2004; Kuhlman et al., 2007). Cells were grown for 48 hours, with inoculation of fresh medium with 1/50 of the previous culture every 16 hours. All bacterial cultures were then pelleted and rinsed twice in ice-cold TBS, resuspended in 6 ml binding buffer (10 mM Tris (pH 7.3), 1 mM EDTA, 50 mM KCl, 100 $ \mu $g/mL bovine serum albumin, 5\% v/v glycerol). Each bacterial culture was then split into two parts, one of which was incubated 30 min further in presence of 1\% formaldehyde, then treated for 5 min with 0.5 M glycine, repelleted and resuspended in 3 ml binding buffer. Mechanical lysis of the cells and moderate chromosome fragmentation were achieved simultaneously by sonication (3 pulses of 30 s separated by 30 s in thawing ice). After centrifugation and elimination of pellets containing remnants and unbroken cells, LacI antibody (Rabbit anti-Lac I antibody 600-401-B04 Rockland Immunochemicals Inc., P.O. BOX 326, Gilbertsville, Pennsylvania, USA) was incubated with an excess of bacterial lysate supernatant (500 $ \mu $l), overnight on a rotating wheel at 4$^\circ$C. 50 $ \mu $l Protein A coupled magnetic beads (Invitrogen Dynal AS, Oslo, Norway) where then added and incubated for 4 hours at 4$^\circ$C. Beads were collected using the Dynal magnet. The first supernatant was removed and stored for control PCR and the pellet was rinsed 5 times using binding buffer and transferred to new tubes. As indicated in Fig.2, the pellets were then washed for 24 or 72 hours with either 0.35 M or 0.5 M NaCl. The presence of DNA fragments in the pellets was then assessed by PCR (28 cycles: denaturation 30 s 95$^\circ$C, annealing 30 s 55$^\circ$C and elongation 30 s 72$^\circ$C) using primers located in the N-acetylglutamate kinase (ArgB) gene unrelated to \textit{lac} (\textit{ArgB}-F: 5'-AGGTTTGTTTCTCGGTGACG-3'; \textit{ArgB}-R: 5'-GTTGCCCTTCGTCTGTTACG-3' yielding a 167 bp-long amplification fragment) or to the lactose promoter P\textit{lac} (P\textit{lac}-F: 5'-TCCGGCTCGTATGTTGTGTG-3'; P\textit{lac}-R 5'-AGGCGATTAAGTTGGGTAACG-3', yielding a 142 bp-long amplification fragment). The same results were obtained in 3 independent experiments starting from fresh bacteria. The molecular weight marker is the 100 bp GeneRuler from Fermentas.

\subsection{$ \beta $-galactosidase assays}
Bulk expression assays were performed using 4-methylumbelliferone-$ \beta $-D-galactopyranoside (MUG) as a substrate, as described in (Vidal-Aroca et al., 2006). For in situ $ \beta $-Galactosidase staining, cells were incubated with 1 mg/ml 5-bromo-4-chloro-3-indolyl-$ \beta $-D-galactopyranoside (X-gal), 2 mM \ce{MgCl2}, 5 mM \ce{K3Fe(CN)6}, 5 mM \ce{K4Fe(CN)6} in phosphate buffered saline, until appearance of blue staining. They were then rinsed and fixed for 1 min in 1\% glutaraldehyde, 1mM \ce{MgCl2}.

\section{Discussion}
The present interpretations of \textit{lac} features are based on the redefinition of the single event early postulated in 1957 by Novick and Weiner. This event cannot be the synthesis of a single molecule of permease as supposed by Novick and Weiner, because of the necessity of a threshold of permeases. The single event can no more be a single transcription round of the \textit{lac} gene, since a steady state basal expression has been evidenced. Hence, alternatively, the single event could be the complete release of the \textit{lac} repressor, in line with the simplistic mode of action of LacI described in the initial reports, but which now appears unusual under the current assumption that interactions are highly dynamic in the cell. Static interactions, which are obvious for multimolecular complexes, are rediscovered in eukaryotic cells (Hemmerich et al., 2011) and are completely conceivable for prokaryotic repressor complexes. A simple way to prolong molecular association in the cell is the hierarchical binding of capping molecules to preformed molecular assembly, leading to the molecular trapping of nucleating components in absence of active dismantling (Michel, 2011). While in quasi-equilibrium approaches, the interactions between LacI and P\textit{lac} are considered as faster than the \textit{lac} regulatory circuit, in the present model they are slower. In presence of low inducer concentration, the mechanisms described in this study are expected to  maintain the presence of uninduced cells in the population in several ways: by strongly reducing the number of bacteria with open P\textit{lac}, by erasing the memory of previous bursts of expression, and by adjusting the probability of switching \textit{lac} on in a single burst. The actual mechanism of \textit{lac} repression by conjugation of several operators remains controversial, but experiments of deletion of individual operators, as well as of mutation of LacI making it non-tetramerizable, strongly decreased \textit{lac} repression, pointing out the importance of a supramolecular repression complex. In absence of inducer, a single LacI tetramer would be locked forever in a large molecular complex including a stabilized DNA hairpin with two operators (Brenowitz et al., 1991). Of course, such a stability cannot be recovered using reporter plasmids and tagged modified non-tetramerizable artificial LacI constructs, for which graded quasi-equilibrium interpretations can hold, contrary to endogenous \textit{lac} components. Owing to its unicity, LacI dissociation would work as a turning point of the \textit{lac} system, disconnecting the preceding and the following processes. Briefly, those occurring before complete LacI dissociation explain populational behaviors and generate a first filter for low inducer concentration, while those occurring after dissociation can still contribute to the all-or-nothing phenotypes. To describe this situation, before LacI dissociation, only one LacI tetramer has to be taken into consideration: the one precisely bound to P\textit{lac}; whereas after dissociation, all LacI tetramers present in the cell should be incorporated in the treatment, since at low inducer concentration, certain can remain inducer-free. The shape of the population curves could directly reflect the probability of occurrence of a single event, that itself depends on the achievement of a series of previous events. Under this assumption, the sigmoidal \textit{lac} induction curves are purely populational effects skipping the single cell level. In this alternative picture, the inherently unpredictable all-or-nothing behavior is unrelated to the population response because of the uncoupling action of the fast positive feedback taking place in individual bacteria. A single stochastic event can generate smooth population behaviors, in the same manner that unpredictable microscopic association and dissociation jumps are converted into smooth exponential curves in surface plasmon resonance experiments. The figure 2 of the article of Novick and Weiner could illustrate particularly well the fact that a purely mathematical notion: the probability of occurrence (continuous and ranging from 0 to 1), of a binary event (discontinuous, with two discrete values 0 or 1), directly translates into the concrete partition of all-or-none phenotypes in the bacterial culture. Hence, \textit{lac} is a typical system for which the population behavior is well defined whereas the state of each cell is indeterminate. In the same way, the result of the molecular race following de-repression is also dictated by chance only, but can be captured probabilistically, that is to say deterministically, at the population level. The example of \textit{lac} shows that populational studies and classical deterministic treatments can be fully appropriate for shedding light on microscopic phenomena through materializing probability distributions. The mechanisms proposed here to give rise to the coexistence of uninduced and induced phenotypes in the population are transient, mathematically speaking, but can in practice lead to durable bimodality. (\textbf{i}) A random walk to an absorbing state is inherently transient but can be astonishingly long for heterogeneous random walks with predominant backward transitions as obvious in Eq.(1), and its individual arrivals are widely dispersed in time. For instance, for a succession of $ n $ identical transitions and when $ k^{-}>>k^{+} $, the mean time and standard deviation reach the same geometric function of the number of steps $ \sigma(T)=\left \langle T \right \rangle \sim (1/k^{+})(k^{-}/k^{+})^{n-1} $. (\textbf{ii}) The single race following derepression can, for appropriate parameter ranges, provide a simple general mechanism to limit cellular induction to a small fraction of the population. One cannot exclude that highly desynchronized single events explosively amplified by positive feedback in every cell contribute for a significant fraction of the observed all-or-nothing behavior of \textit{lac}. There is some confusion about \textit{lac} bistability, sometimes put forward to describe the bimodal distribution of bacteria. In fact, bistability is not necessary to explain certain features of \textit{lac} and the binary individual phenotypes can be unrelated to sigmoidal responses to inducer concentration. Examples of mechanisms that can underlie different possible situations in which single cell and population behaviours are disconnected, are listed in Appendix A2. In these examples, the principle of ligand-induced de-repression is retained (i.e. inhibition by a DNA-binding repressor, itself suppressed by an inducer). In fact, \textit{lac} is supposed to be bistable not only to explain the prolonged bimodal partition of the population, but most importantly to explain the existence of an induction threshold preventing the generalized induction of bacteria at low inducer concentration and solving the apparent paradox that \textit{lac} expression is necessary for \textit{lac} induction. These characteristics are to some extent also obtained with the kinetic mechanisms described here. The basal expression of permease allows bacteria to sense the presence of lactose in their environment and a basal level of $ \beta $-galactosidase is also required to isomerize lactose into allolactose. But these basal levels should be low enough to avoid priming the positive feedback when lactose concentration is insufficient to justify the metabolic investment for the bacteria to synthesize \textit{lac} machineries. From a physiological perspective and if gratuitous inducers have some relevance, this mechanism would be of great evolutionary interest since it offers different advantages: (\textbf{i}) At the level of a single cell, it allows that the decision to use or not lactose is unequivocal, either minimal or maximal. (\textbf{ii}) At the level of the population, particularly when the substrate concentration is low and uncertain, it allows that only a fraction of the cells invests into this expensive specialization reducing the division rate.\\
\newline
	Beyond its apparent simplicity, the \textit{lac} system remains largely enigmatic. For complex biochemical systems, inaccurate models can be unintentionally decorated by misleading experiments. Even when reduced to its basic components, reliable experimental results can be interpreted through several ways, leading to strikingly different models. In the present case, S-shaped dose-response curves can be interpreted as well using quasi-equilibrium cooperativity (Oehler et al., 2006), with similar outputs for single cells and populations, or as the distribution of a transient process, for which single cell and populations behaviours can be disconnected (Appendix A2). The existence of \textit{lac} bistability also remains to be confirmed (Savageau, 2011). Alternative mechanisms possibly contributing to \textit{lac} behaviors and justifying the tetrameric structure of LacI are added here to the debate, but the saga of the lactose operon is likely to remain far from complete.

\end{multicols}
\newpage
\section{Appendices}
\subsection{Appendix 1: Random walk modeling of the chain of events leading to micro-irreversible LacI dissociation}
The 15 transitions represented in Fig.3b are not elementary events but are themselves composite reactions including a lot of micro-substeps (individual chemical bond breakages and so on), so that their designation as Poisson transitions is only approximate, particularly for pseudo-first-order reactions. In addition to the minimal chain drawn in Fig.3b, P\textit{lac} clearance is also conditioned by other upstream phenomena necessary to convey the inducer close to LacO and many other interfering parameters including negative effect of \textit{lac} induction on the growth rate, molecules dilution or crowding, diffusion and translocation, bacterial background, metabolism, accessory factors present in the repression complex etc. It can however be proposed that the most elementary biochemical transitions, the "atoms of dynamics", refractory to experimental measurements, are naturally and necessarily distributed according to an exponential law, because this distribution is the only one that is fully memoryless. In real biochemical networks, every transition is conditioned by specific activation energies, so that the rates can be different. The probabilities of the forward and backward jumps starting from any state $ x_{i} $, are not 1/2 and 1/2 as for isotropic random walks, but $ k_i^+/(k_i^-+k_i^+) $ and $ k_i^-/(k_i^-+k_i^+) $ respectively. Different approaches can be used to calculate the probability of completion of such chains, including a discrete random walk approach, in which $ x_{0} $ is a reflecting wall and $ x_{n} $ is an absorbing one, and a time continuous Markovian approach using a set of linear differential equations with an absorbing final state $ (k_n^- = 0) $ and solving Eq.(A1). The system can be described with the same rule for every state $ i $ such that $ 1 < i < n-1 $, as follows\\ 
\newline
\newline
$ \begin{pmatrix}
\dot{x}_{0}\\ 
\hphantom{\vdots} \\
..\\ 
\hphantom{\vdots} \\
\dot{x}_{i}\\ 
\hphantom{\vdots} \\
..\\ 
\hphantom{\vdots} \\
\dot{x}_{n-1}\\ 
\hphantom{\vdots} \\
\dot{x}_{n}\\ 
\end{pmatrix}
=
\begin{pmatrix}
-k^{+}_{0} & k^{-}_{1} & ... & 0 & 0 & 0 & ... & 0 & 0 & 0 \\ 
... & ... & ... & ... & ... & ... & ... & ... & ... & ... \\
0 & 0 & ... & k^{+}_{i-1} & -(k^{-}_{i}+k^{+}_{i}) & k^{-}_{i+1} & ... & 0 & 0 & 0 \\ 
... & ... & ... & ... & ... & ... & ... & ... & ... & ... \\
0 & 0 & ... & 0 & 0 & 0 & ... & k^{+}_{n-2} & -(k^{-}_{n-1}+k^{+}_{n-1}) & 0 \\
0 & 0 & ... & 0 & 0 & 0 & ... & 0 & k^{+}_{n-1} & 0 \\
\end{pmatrix}
\begin{pmatrix}
x_{0}\\ 
\hphantom{\vdots} \\
..\\ 
\hphantom{\vdots} \\
x_{i}\\ 
\hphantom{\vdots} \\
..\\ 
\hphantom{\vdots} \\
x_{n-1}\\ 
\hphantom{\vdots} \\
x_{n}\\ 
\end{pmatrix} $ \\ 

\begin{flushright}
(A1)\\
\end{flushright}
\textbf{Correspondences between discrete and time continuous random walk approaches}\\
\\
If using a discrete random walk approach, Eq.(4a) is the complement to one of the probability to remain confined in the $ n $-1 first positions. This probability simplifies for large $ N $ and $ n $ and for a moderate ratio $ N/n^2 $ as described in (Ruelle, 2006). In this case, $ \theta $ is the ratio of the number of jumps over the square of the number of steps in the chain (Eq.(4b)), that is itself the mean number of jumps required to complete the chain, according to the well established random diffusion rate.\\
For the continuous time differential approach with a single rate $ k $, the probability of state $ n-1 $ is the probability density of state $ n $. Finally it is even more useful to make $ \theta $ independent of the number of jumps and of steps, because these microscopic values are generally unknown in real chains of events. The equivalence between the discrete and time continuous approaches can be derived as follows: The mean time of a single jump $ \tau $, such that $ t = N \tau $, is $ \tau = (n+1)/2nk $ (considering that the waiting times are 1/$ k $ for the state $ x_0 $ and 1/2$ k $ for all the other positions) and given the mean completion time of the whole isotropic chain (Eq.(3)), $ N/n^2 $ is equivalent to $ t/\left \langle T \right \rangle $. 

\subsection{Appendix 2: Table A1 }
Single cell binarity and population sigmoidicity are not necessarily related. The ligand-dependent transcriptional probability functions shown in absence of self-regulated circuits, illustrate the possible disconnection between sigmoidicity and bimodality.\\
\newline
\textbf{Left column}. Single cell (symbolized by disks) and population (curves) responses to inducer. Greyscale (graded) or bitmap (all-or-nothing) expression intensities can be obtained through fast equilibrium, in panels (a) and (c), or single event-primed switches in panels (b) and (d), respectively. For the single event mechanisms, the system is freezed at a given time after inducer addition. The individual behaviors are not necessarily related to the apparent population induction shape, and bimodality does not always require population sigmoidicity. In panels (a) and (b), the populational responses are non-sigmoidal while in panels (c) and (d), they are sigmoidal, also said to be ultrasensitive.\\
\newline
\textbf{Right column}. Examples of equations underlying the different situations shown in the left column. Non-zero values of $ F $ are possible for [$ A $] = 0 in the fast equilibrium mechanisms. The equivalence between population and single cell behaviors in the fast equilibrium mechanisms, relies on the ergodic equivalence between space averaging and time averaging respectively.\\ 
\newline
(\textbf{a}) The repressor ($ R $) interacts with DNA through short association/dissociation cycles and the inducer ($ A $) determines the fraction of time in which the promoter ($ Prom $) is occupied. $ K_R $ is the binding equilibrium constant between $ R $ and $ Prom $ and $ K_A $ the binding constant of the inducer with $ R $.\\ 
\newline
(\textbf{b}) The stochastic dissociation of the repressor is a very single stochastic event triggered by the inducer and corresponding to the release of monomeric repressor $ R $ from DNA. This mechanism is less likely to occur than the chain of events postulated in (d), because of the inevitable risk of unwanted priming in fluctuating biological systems.\\ 
\newline
(\textbf{c})  As in panel (a), the inducer equilibrates rapidly with the repressor and the repressor rapidly equilibrates with DNA, so that equilibrium cooperativity can apply to the saturation of $ R $ by the inducer, as well as to the saturation of $ Prom $ by $ R $. The formulations shown use the MWC model of cooperativity which necessitates few parameters and has already been used for LacI. Three cases can be envisioned.\\ 
•	$ c_1 $ Cooperative fixation of the inducer to $ R $. $ C_R $ corresponds to the repressor conformational equilibrium constant.\\ 
•	$ c_2 $ Cooperativity concerns the fixation of $ R $ to the two operators of \textit{Prom}. $ C_L $ is the $ Prom $-looping constant and $ P $ is the fraction of unliganded $ R $ supposed to be bound non-cooperatively by $ A $.\\
A mix of $ c_1 $ and $ c_2 $, with involvement of both $ C_R $ and $ C_L $, is also possible (not formulated). \\ 
\newline
(\textbf{d}) Model retained in the present study. The induced phenotype appears once the final event of a chain conditioned by inducer addition, occurs. The example of equation shown corresponds to a chain of two consecutive events, in which the first one is reversible and the two forward transitions depend on the inducer (Fig.A1).\\ 

\begin{center}
\includegraphics[width=14.2cm]{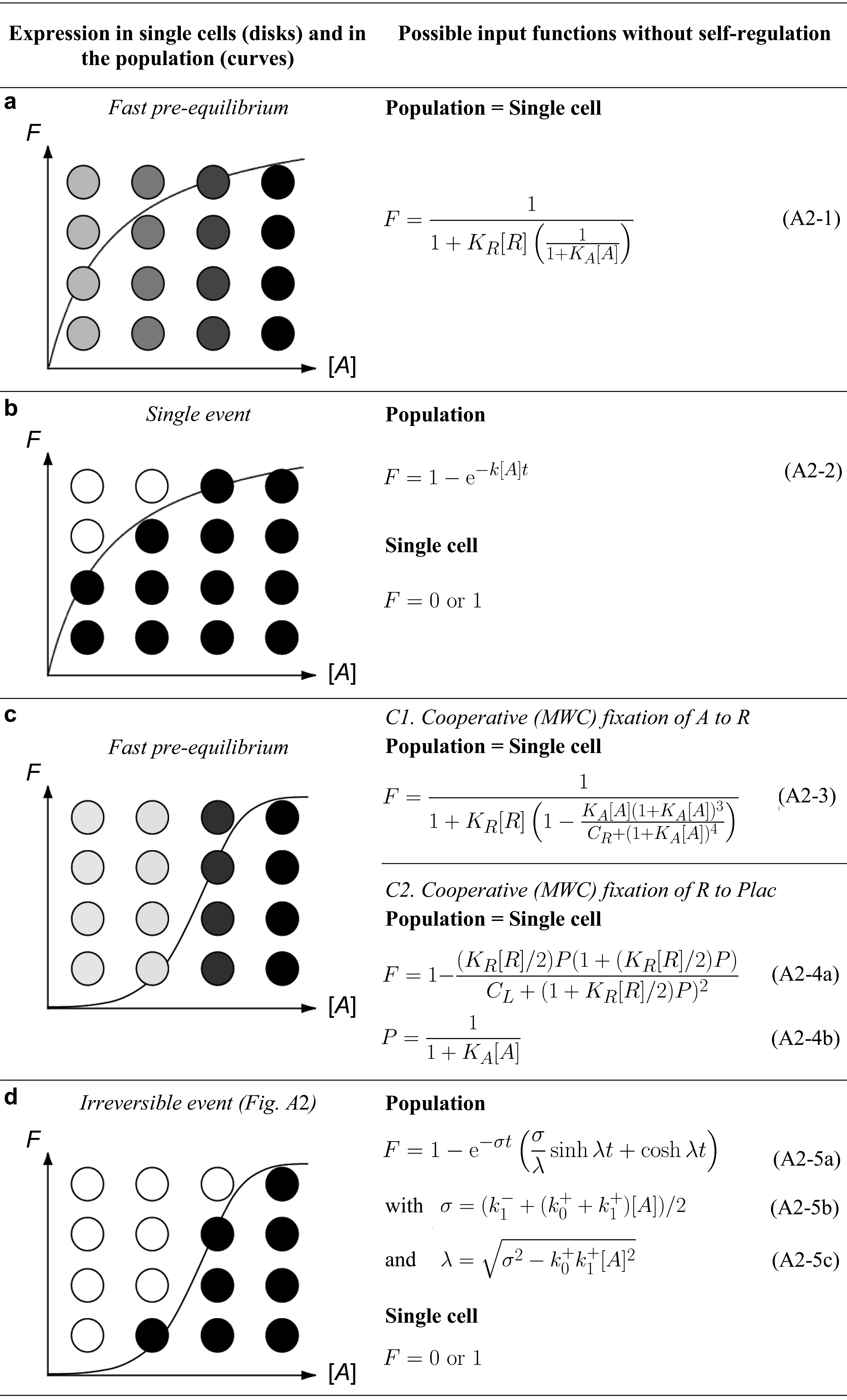} \\
\end{center}

\begin{multicols}{2}
\begin{center}
\includegraphics[width=8cm]{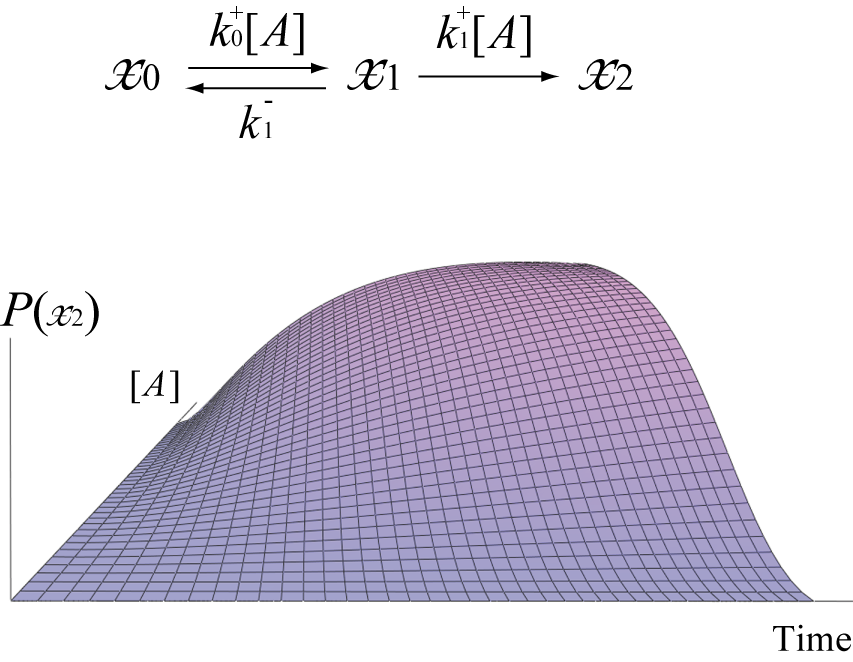} \\
\end{center}

\textbf{Figure A2}. \begin{small} Time- and concentration-dependent completion of a two-step chain with two pseudo-first-order forward rates. The probability of state $ x_2 $ is a function of the combination of time after inducer addition and of inducer concentration. 3D plot drawn to Eq.(A2-5) with $ k_1^-/k_0^+ = k_1^-/k_1^+ = 3 $ and $ P(x_0)=1 $ at $ t_0=0 $ \end{small}. \\
\newline
\textbf{Acknowledgements} I thank the Carlos Blanco's lab for giving the \textit{E. coli} K12 cells and synthetic media; Philippe Ruelle and Jean-Christophe Breton for discussions helpful for building Eq.(4); Catherine Martin-Outerovitch and Henri Wr\'oblewski for their help in bacterial staining and microscopy and the anonymous reviewers for very useful suggestions.
\newline
\section{References}
 \begin{small}
Balaeff, A., Mahadevan, L., Schulten, K. 2004. Structural basis for cooperative DNA binding by CAP and lac repressor. Structure 12, 123-132.\\
Barkley, M.D., Riggs, A.D., Jobe, A., Burgeois, S. 1975. Interaction of effecting ligands with lac repressor and repressor-operator complex. Biochemistry 14, 1700-1712.\\
Beard, D.A. 2011. Simulation of cellular biochemical system kinetics. Wiley Interdiscip. Rev. Syst. Biol. Med. 3, 136-146.\\
Benzer, S. 1953. Induced synthesis of enzymes in bacteria analyzed at the cellular level. Biochim. Biophys. Acta 11, 383-395.\\
Berg, O.G., Blomberg, C. 1977. Mass action relations in vivo with application to the lac operon. J. Theor. Biol. 67, 523-533.\\
Bolouri, H., Davidson, E.H. 2003. Transcriptional regulatory cascades in development: initial rates, not steady state, determine network kinetics. Proc. Natl. Acad. Sci. U.S.A. 100, 9371-9376.\\
Brenowitz, M., Pickar, A., Jamison, E. 1991. Stability of a Lac repressor mediated “looped complex”. Biochemistry 30, 5986-5998.\\
Chen, J., Alberti, S., Matthews, K.S. 1994. Wild-type operator binding and altered cooperativity for inducer binding of lac repressor dimer mutant R3. J. Biol. Chem. 269, 12482-12487.\\
Chen, J., Matthews, K.S. 1994. Subunit dissociation affects DNA binding in a dimeric lac repressor produced by C-terminal deletion. Biochemistry 33, 8728-8735.\\
Cherry, J.L., Adler, F.R. 2000. How to make a biological switch. J. Theor. Biol. 203, 117-133.\\
Choi, P.J., Cai, L., Frieda, K., Xie, X.S. 2008. A stochastic single-molecule event triggers phenotype switching of a bacterial cell. Science 322, 442-446.\\
Choi, P.J., Xie, X.S., Shakhnovich, E.I. 2010. Stochastic switching in gene networks can occur by a single-molecule event or many molecular steps. J. Mol. Biol. 396, 230-244.\\
Chung, J.D., Stephanopoulos, G. 1996. On physiological multiplicity and population heterogeneity of biological systems. Chem. Eng. Sci. 51, 1509-1521.\\
Daber, R., Sharp, K., Lewis, M. 2009. One is not enough. J. Mol. Biol. 392, 1133-1144.\\
D\'iaz-Hern\'andez, O., Santill\'an, M. 2010. Bistable behavior of the lac operon in E. coli when induced with a mixture of lactose and TMG. Front Physiol 1, 22.\\
Doan, T., Mendez, A., Detwiler, P.B., Chen, J., Rieke, F. 2006. Multiple phosphorylation sites confer reproducibility of the rod’s single-photon responses. Science 313, 530-533.\\
Dunaway, M., Manly, S.P., Matthews, K.S. 1980. Model for lactose repressor protein and its interaction with ligands. Proc. Natl. Acad. Sci. U.S.A. 77, 7181-7185.\\
Elf, J. Li, G.W., Xie, X.S. 2007. Probing transcription factor dynamics at the single-molecule level in a living cell. Science 316, 1191-1194.\\
Ferrell, J.E. Jr. 2012. Bistability, bifurcations, and Waddington’s epigenetic landscape. Curr. Biol. 22, R458-R466.\\
Fried, M.G., Hudson, J.M. 1996. DNA looping and lac repressor-CAP interaction. Science 274, 1930–1931; author reply 193 1-1932.\\
Garcia, H.G., Phillips, R. 2011. Quantitative dissection of the simple repression input-output function. Proc. Natl. Acad. Sci. U.S.A. 108, 12173-12178.\\
Golding, I., Paulsson, J., Zawilski, S.M., Cox, E.C. 2005. Real-time kinetics of gene activity in individual bacteria. Cell 123, 1025-1036.\\
Becker, N.A., Kahn, J.D., Maher III, L.J. 2007. Effects of nucleoid proteins on DNA repression loop formation in \textit{Escherichia coli}. Nucl. Acids. Res. 35, 3988-4000.\\
Grainger, D.C., Hurd, D., Goldberg, M.D., Busby, S.J.W. 2006. Association of Nucleoid Proteins with Coding and Non-Coding Segments of the Escherichia Coli Genome. Nucl. Acids Res. 34, 4642-4652.\\
Hemmerich, P., Schmiedeberg, L., Diekmann, S. 2011. Dynamic as well as stable protein interactions contribute to genome function and maintenance. Chromosome Res. 19, 131-151.
Jacob, F. 2011. The birth of the operon. Science 332, 767.\\
Jacob, F., Monod, J. 1961. Genetic regulatory mechanisms in the synthesis of proteins. J. Mol. Biol. 3, 318-356.\\
Jacob, F., Monod, J. 1963. Genetic repression, allosteric inhibition, and cellular differentiation. In Cytodifferential and Macromolecular Synthesis, Ed Locke M. New York Academic Press 30-64.\\
Koshland, D.E., Jr, N\'{e}methy, G., Filmer, D. 1966. Comparison of experimental binding data and theoretical models in proteins containing subunits. Biochemistry 5, 365-385.\\
Kuhlman, T., Zhang, Z., Saier, M.H. Jr, Hwa, T. 2007. Combinatorial transcriptional control of the lactose operon of Escherichia coli. Proc. Natl. Acad. Sci. U.S.A. 104, 6043-6048.\\
Lee, J., Goldfarb, A. 1991. lac repressor acts by modifying the initial transcribing complex so that it cannot leave the promoter. Cell 66, 793-798.\\
Lewis, M. 2005. The lac repressor. C. R. Biol. 328, 521-548.\\
Michel, D. 2009. Fine tuning gene expression through short DNA-protein binding cycles. Biochimie 91, 933-941.\\
Michel, D. 2010. How transcription factors can adjust the gene expression floodgates. Prog. Biophys. Mol. Biol. 102, 16-37.\\
Michel, D. 2011. Basic statistical recipes for the emergence of biochemical discernment. Prog. Biophys. Mol. Biol. 106, 498-516.\\
Moazed, D. 2011. Mechanisms for the inheritance of chromatin states. Cell 146, 510-518.\\
Nicol-Beno\^{i}t, F., Le-Goff, P., Le-Dr\'{e}an, Y., Demay, F., Pakdel, F., Flouriot, G., Michel, D. 2012. Epigenetic memories: structural marks or active circuits? Cell. Mol. Life Sci. 69, 2189-2203.\\
Novick, A., Weiner, M. 1957. Enzyme induction as an all-or-none phenomenon. Proc. Natl. Acad. Sci. U.S.A. 43, 553-566.\\
O'Gorman, R.B., Rosenberg, J.M., Kallai, O.B., Dickerson, R.E., Itakura, K., Riggs, A.D., Matthews, K.S. 1980. Equilibrium binding of inducer to lac repressor.operator DNA complex. J. Biol. Chem. 255, 10107-10114.\\
Oehler, S., Alberti, S., M\"{u}ller-Hill, B. 2006. Induction of the lac promoter in the absence of DNA loops and the stoichiometry of induction. Nucl. Acids Res. 34, 606-612.\\
Oehler, S., Eismann, E.R., Kr\"{a}mer, H., M\"{u}ller-Hill, B. 1990. The three operators of the lac operon cooperate in repression. EMBO J. 9, 973-979.\\
Ozbudak, E.M., Thattai, M., Lim, H.N., Shraiman, B.I., Van Oudenaarden, A. 2004. Multistability in the lactose utilization network of Escherichia coli. Nature 427, 737-740.\\
Ruelle, P. 2006. Lectures notes on random walks PHYS 2122\\(http://www.fyma.ucl.ac.be/data/pdf/random-walks-v4.pdf)\\ 
Sadler, J.R., Sasmor, H., Betz, J.L. 1983. A perfectly symmetric lac operator binds the lac repressor very tightly. Proc. Natl. Acad. Sci. U.S.A. 80, 6785-6789.\\
Sanchez, A., Osborne, M.L., Friedman, L.J., Kondev, J., Gelles, J. 2011. Mechanism of transcriptional repression at a bacterial promoter by analysis of single molecules. EMBO J. 30, 3940-3946.\\
Savageau, M.A. 2011. Design of the lac gene circuit revisited. Math. Biosci. 231, 19-38.\\
Setty, Y., Mayo, A.E., Surette, M.G., Alon, U. 2003. Detailed map of a cis-regulatory input function. Proc. Natl. Acad. Sci. U.S.A. 100, 7702-7707.\\
Sharp, K.A. 2011. Allostery in the lac operon: population selection or induced dissociation? Biophys. Chem. 159, 66-72.\\
Simons, A., Tils, D., von Wilcken-Bergmann, B., M\"{u}ller-Hill, B. 1984. Possible ideal lac operator: Escherichia coli lac operator-like sequences from eukaryotic genomes lack the central G-X-C pair. Proc. Natl. Acad. Sci. U.S.A. 81, 1624-1628.\\
Sobie, E.A. 2011. Bistability in biochemical signaling models. Sci Signal 4, tr10.\\
Thomas, R. 1998. Laws for the dynamics of regulatory networks. Int. J. Dev. Biol. 42, 479-485.\\
Tolker-Nielsen, T., Holmstrøm, K., Boe, L., Molin, S. 1998. Non-genetic population heterogeneity studied by in situ polymerase chain reaction. Mol. Microbiol. 27, 1099-1105.\\
Veliz-Cuba, A., Stigler, B. 2011. Boolean models can explain bistability in the lac operon. J. Comput. Biol. 18, 783-794.\\
Vidal-Aroca, F., Giannattasio, M., Brunelli, E., Vezzoli, A., Plevani, P., Muzi-Falconi, M., Bertoni, G. 2006. One-step high-throughput assay for quantitative detection of beta-galactosidase activity in intact gram-negative bacteria, yeast, and mammalian cells. BioTechniques 40, 433-434.\\
Vilar, J.M.G., Guet, C.C., Leibler, S. 2003. Modeling network dynamics: the lac operon, a case study. J. Cell Biol. 161, 471-476.\\
Wilson, C.J., Zhan, H., Swint-Kruse, L., Matthews, K.S. 2007. The lactose repressor system: paradigms for regulation, allosteric behavior and protein folding. Cell. Mol. Life Sci. 64, 3-16.\\
Xu, M., Long, C., Chen, X., Huang, C., Chen, S., Zhu, B. 2010. Partitioning of histone H3-H4 tetramers during DNA replication–dependent chromatin assembly. Science 328, 94-98.
Yagil, G., Yagil, E. 1971. On the relation between effector concentration and the rate of induced enzyme synthesis. Biophys. J. 11, 11-27.\\
Yildirim, N., Kazanci, C. 2011. Deterministic and stochastic simulation and analysis of biochemical reaction networks: The lactose operon example. Methods Enz. 487, 371-395.\\
Zhan, H., Camargo, M., Matthews, K.S. 2010. Positions 94-98 of the lactose repressor N-subdomain monomer-monomer interface are critical for allosteric communication. Biochemistry 49, 8636-8645.\\
 \end{small}
\end{multicols}
\end{document}